\documentclass[12pt]{article}
\usepackage{amsthm, amsmath, amssymb}
\usepackage{latexsym,amsmath,amsfonts,amsthm,amssymb}

\newtheorem{theorem}{Theorem}[section]
\usepackage{tikz}
\usepackage[dvips]{epsfig}
\usepackage{enumerate}
\usepackage{bbm}
\usepackage{graphics}
\usepackage[font={footnotesize},margin=1cm]{caption}

\usepackage[normalem]{ulem}
\usepackage[breaklinks=false,pdfpagemode=None,pdfview=FitH,pdfstartview=FitH,citebordercolor={0 0 1},linkbordercolor={0 0 1},urlbordercolor={0 0 1},pagebordercolor={0 0 1},pdfborder={0 0 1}]{hyperref}

\usetikzlibrary{arrows,shapes}
\usetikzlibrary{decorations.pathmorphing}

\newcommand{\veps}{\varepsilon}

\newcommand{\calG}{{\mathcal{G}}}
\newcommand{\calH}{{\mathcal{H}}}

\newcommand{\calP}{{\mathcal{P}}}

\newcommand{\calS}{{\mathcal{S}}}

\newcommand{\tc}{\textcolor}

\numberwithin{equation}{section}

\newtheorem{corollary}[theorem]{Corollary}
\newtheorem{lemma}[theorem]{Lemma}
\newtheorem{definition}[theorem]{Definition}
\newtheorem{proposition}[theorem]{Proposition}
\title  {
        Level Spacing for Non-Monotone Anderson Models
                 }

\author{
John Z.\ Imbrie\footnote{
This research was conducted in part while
the author was visiting the Institute for Advanced Study in Princeton, supported by The Fund for Math and The Ellentuck Fund.
}
\\Department of Mathematics,
University of Virginia \\
Charlottesville, VA 22904-4137, USA
\\ {\tt imbrie@virginia.edu}
\\
\\Rajinder Mavi
\\Department of Mathematics,
University of Virginia \\
Charlottesville, VA 22904-4137, USA
\\ {\tt rsm8y@virginia.edu}
}
\date{}
\begin{document}
\maketitle
\begin{abstract}
We prove localization and probabilistic bounds on the minimum level spacing for a random block Anderson model without monotonicity. Using a sequence of narrowing energy windows and associated Schur complements, we obtain detailed probabilistic information about the microscopic structure of energy levels of the Hamiltonian, as well as the support and decay of eigenfunctions.
\end{abstract}
\tableofcontents
\section{Introduction}\label{sec:introduction}
\subsection{Background}\label{sect:background}
We present a new method for proving separation of eigenvalues for disordered quantum mechanical Hamiltonians on the lattice, in the strong disorder or weak hopping regime. The traditional approach to this question is based on Minami's method \cite{Minami1996}. When available, it provides a bound on the probability of two or more eigenvalues in an interval, and thus gives probabilistic control on the separation of eigenvalues. Variants of the Minami estimate have been an  important \tc{black}{component} of proofs of Poisson statistics for eigenvalues in an interval \cite{Minami1996, Nakano2006, Killip2006, Graf2006, Bellissard2007, Combes2009, Germinet2014, Germinet2012}. They are  important for understanding the behavior of the ac conductivity \cite{Klein2007}. They \tc{black}{also} lead to minimum level-spacing estimates and simplicity of the spectrum, provided the probability distribution of the potential is H\"older continuous with index $\alpha > \tfrac{1}{2}$ \cite{Klein2006,Combes2009}.

Obtaining Minami-type estimates has been problematic for so-called `non-monotone' models, in which the variation of the single-site potential with the random variables may change in sign from point to point in space. While there are a number of results on localization for non-monotone models \cite{Klopp1995, Veselic2002, Combes2002, Veselic2010, Veselic2010a, Elgart2010,  Elgart2011, Kruger2011, Cao2012, Elgart2014}, there are only a couple of multi-level results that we are aware of. 
Simplicity of the spectrum in infinite volume is established in \cite{Naboko2011}.
In \cite{Tautenhahn2013}, a Minami estimate is proven for some non-monotone models that admit a transformation to a monotone situation. In \cite{Elgart2013} an 
$N$-level Minami estimate is proven for a class of random block models, in which a $k\times k$ random block Hamiltonian sits at each site of the lattice, and the blocks are coupled through a deterministic hopping matrix. The randomness must be sufficiently rich to ensure the success of the method. Of particular interest is the case of $2\times2$ blocks of the form $ \Big( \begin{array}{cc}
u &v \\v & -u  \end{array} \Big)$;
\tc{black}{Minami estimates were obtained in}
 \cite{Elgart2013} in the situation in which both $u$ and $v$ are $\beta$-regular random variables. Still, this leaves out the case where $v$ is deterministic, which may be called a fully diluted model (meaning that there is only one degree of freedom in the randomness for each block). 

In this work we consider a fully diluted random block Hamiltonian with $2\times2$ block $ \Big( \begin{array}{cc}
u &1 \\1 & -u  \end{array} \Big)$. This is a model originally proposed by Spencer; localization and H\"older continuity of the integrated density of states are proven in \cite{Elgart2014}. We prove a probabilistic statement on minimum eigenvalue spacing. Specifically, we give a bound showing that the probability that the minimum eigenvalue separation is less than $\delta$ tends to zero with $\delta$. (See Theorem \ref{thm:2.3} below). Our interest in this problem stems from recent work \cite{Imbrie2014} in which a minimum eigenvalue separation condition (called limited level attraction) was used as an assumption in a proof of many-body localization for a one-dimensional spin chain. Although the separation condition remains unproven in the many-body context, one may perhaps gain some insight into the problem by replacing the tensor product of $2\times2$ spin spaces with a direct sum, and this leads to the random block Hamiltonian considered in this paper. The dilution of randomness is not as extreme as in the many-body case, but it remains an important issue nevertheless.

The method should be of interest in its own right. We use a sequence of Schur complements to reduce the Hilbert space to subspaces of smaller and smaller dimension. At the same time, we shrink the width of the energy interval under consideration, and eventually only one eigenvalue remains. The subspaces are structured in physical space as a set of extended blocks whose (renormalized) spectrum is in a small neighborhood of some energy $E$. The Schur complement formula produces effective Hamiltonians for the blocks exhibiting exponential decay in the distance between blocks. The blocks themselves become more and more dilute as the procedure proceeds, so the distance between blocks grows, and hence inter-block matrix elements tend rapidly to zero. The key probabilistic input involves estimating the probability that a block has spectrum close to $E$, and bounding the probability that a block has a very small level spacing. The determinant and the discriminant of the Schur complement matrix for a block is a polynomial in the random variables (the degree depends on the size of the block). This ensures a degree of transversality  of  eigenvalues of subsystems (and their differences) -- although not uniformly in the size of the subsystem. We obtain a rather weak bound on the probability of spectrum near $E$ (replacing the usual Wegner estimate \cite{Wegner1981}) and the probability of near-degeneracy (replacing the usual Minami estimate). Although the bound degenerates as the size of the block gets large, large blocks are improbable to begin with, and the trade-off determines the form of our results on density of states and level separation. There is some similarity with the method of \cite{Bourgain2009, Kruger2011}, in which localization is proven for a Hamiltonian depending analytically on the random variables. In those works, the Wegner estimate is replaced with a transversality condition arising from polynomial approximation of determinants (using a version of Cartan Lemma proven in \cite{Bourgain2009}).

As in \cite{Imbrie2014a}, we give a direct construction of eigenfunctions and eigenvalues in terms of convergent graphical expansions that depend on the random variables only within each graph. However, we gain some precision by working near a fixed energy (as advocated by \cite{Chulaevsky2014a}). We have also been influenced by ideas in \cite{Bach1998,Bach2006} (the Feshbach map) and of course by \cite{Anderson1958,Frohlich1983}.

\subsection{Model}\label{sect:model}

We consider the following non-monotone Anderson model. A particle hops in $\Lambda$, a rectangle in $\mathbb{Z}^d$, with each position \tc{black}{$x \in \Lambda$} having two states, + and --. The single-position Hamiltonian is
\begin{equation}
h_x \equiv \left( \begin{array}{cc}
u_x &1 \\1 & -u_x  \end{array} \right),
\label{(1.2)}
\end{equation}
where $\mathbf{u} = \{u_x\}_{x\in \Lambda}$ is a collection of bounded iid random variables with a bounded density. For simplicity, let us assume each $u_x$ has a uniform distribution on [-1,1]. The Hilbert space is $\mathbb{C}^{2n}$, where $n = |\Lambda|$ is the number of elements of $\Lambda$. Let $H_0 = \oplus_{x\in \Lambda}h_x$. The hopping matrix is
\begin{equation}
V = \gamma J \otimes I_2,
\label{(1.3)}
\end{equation}
with $I_2$ denoting the $2\times2$ identity matrix for the internal space, $\gamma$ is a small parameter, and
\begin{equation}
J_{xy} = \begin{cases}
1, &\text{if } |x-y| = 1; \\
0, &\text{otherwise.}
\end{cases}
\label{(1.4)}
\end{equation}
Here $|x| \equiv \sum_{i=1}^d x_i$. Finally, we define the Hamiltonian
\begin{equation}
H = H_0 + V.
\label{(1.5)}
\end{equation}
One could of course consider a much wider class of models. However, this choice will serve to illustrate the method in a simple context exhibiting both non-monotonicity and dilution. That is, the ``bare'' energies $u_x, -u_x$ in (\ref{(1.2)})
move in opposite directions as $u_x$ varies, and there is only one random variable $u_x$ for each $2\times2$ block of $H_0$.

\subsection{\tc{black}{Main} Results}\label{ssec:results}
Our goal is first to prove localization by controlling the density of states and by showing stretched exponential decay of the eigenfunction correlator 
$\mathbb{E} \sum_{\alpha}|\varphi_\alpha(x)\varphi_\alpha(y)|$. Second, we prove probabilistic bounds on level spacing by constructing individual eigenstates and controlling their separation in energy from the others.

We prove several facts about the random block Hamiltonian (\ref{(1.5)}).
\begin{theorem}
\label{thm:1.1}
(Theorem \ref{thm:2.1}). There exists a constant $b>0$ such that for $\gamma$ sufficiently small the following is true. For any $0 < \delta \le \gamma^{1/4}$, any rectangle $\Lambda$, and any interval $I = [E-\delta/2,E+\delta/2]$, let $N(I)$ denote the number of eigenvalues of $H$ in $I$. Then
\begin{equation}
\mathbb{E}\,N(I) \le 4|\Lambda| \exp(-b|\log \gamma|^{2/3}|\log \delta|^{1/3}).
\label{(1.5a)}
\end{equation}
\end{theorem}
Although this bound goes to zero faster than any power of $1/|\log \delta|$, it does not imply 
H\"older continuity of the integrated density of states. Thus it is weaker than the result of 
\cite{Elgart2014}. 
 \begin{theorem}
\label{thm:1.2} (Theorem \ref{thm:2.2}).
Let $\chi = \tfrac{2}{9}$. There exists a constant $\bar{q}>0$ such that for $\gamma$ sufficiently small, the eigenfunction correlator satisfies 
\begin{equation}
\mathbb{E} \sum_\alpha |\varphi_\alpha(y)\varphi_\alpha(z)| \le 4\gamma^{\bar{q}|y-z|^\chi},
\label{(1.5b)}
\end{equation}
for any rectangle $\Lambda$.
\end{theorem}
A similar bound is proven in \cite{Elgart2014}, but with exponential decay.
If we restrict the sum in (\ref{(1.5b)}) to eigenvalues in $I$, we prove a similar bound but with a prefactor tending to zero like the right-hand side of (\ref{(1.5a)}) -- Corollary \ref{cor:2.2'}. \tc{black}{We prove these results partly as a warm-up to the level-spacing problem -- but it is also helpful to understand how strong a result is possible with our method and to make comparisons with the fractional-moment results.}

Finally, we give our main result on minimum level spacing. 
Let $\{E_\alpha\}_{ \alpha = 1,\ldots,2|\Lambda|}$ denote the eigenvalues of $H$.
\begin{theorem}
\label{thm:1.3} (Theorem \ref{thm:2.3}).
There exists a constant $b>0$ such that for $\gamma$ sufficiently small, 
\begin{equation}
P\Big(\min_{\alpha \ne \beta} |E_\alpha  - E_\beta| < \delta\Big) \le |\Lambda| \exp\left(-b|\log \gamma|^{3/4}|\log \delta|^{1/4}\right),
\label{(1.5c)}
\end{equation}
for any rectangle $|\Lambda|$ and any $0<\delta\le \gamma^{1/4}/4$.
\end{theorem}
As this is a result about microscopic level statistics, we are interested in the behavior as $\delta \rightarrow 0$. This bound tends to zero more slowly than (\ref{(1.5a)}), but still faster than any power of
$1/|\log \delta|$.

\subsection{Fundamental Lemma}\label{ssec:fl}
We state here a lemma on Schur complements that will be used often in our analysis.
\begin{lemma}
\textit{(Fundamental Lemma)} Let $K$ be a $(p+q)\times(p+q)$ symmetric matrix in block form, $K = \Big( \begin{array}{cc}
A &B \\C & D  \end{array} \Big)$, with $A$ a $p\times p$ matrix, $D$ a 
$q\times q$ matrix, and $C = B^T$. Assume that $\|(D-E)^{-1}\| \le \veps^{-1},
\|B\| \le \gamma, \|C\| \le \gamma$. Define the Schur complement with respect to 
$\lambda$:
\begin{equation}
F_{\lambda} \equiv A - B(D-\lambda)^{-1}C.
\label{(1.6)}
\end{equation}
Let $\veps$ and $\gamma / \veps$ be small, and $|\lambda-E| \le \veps /2$. Then
\begin{enumerate}
\item{
If $\varphi$ is an eigenvector for $F_\lambda$ with eigenvalue $\lambda$, then
$(\varphi, -(D-\lambda)^{-1}C\varphi)$ is an eigenvector for $K$ with eigenvalue $\lambda$, and all eigenvectors of $K$ with eigenvalue $\lambda$ are of this form.
}
\item{
The spectrum of $K$ in $[E-\veps/2,E+\veps/2]$ is in close agreement with that of $F_E$ in the following sense. If $\lambda_1 \le \lambda_2 \le \ldots \le \lambda_m$ are the eigenvalues of $K$ in $[E-\veps/2,E+\veps/2]$, then there are corresponding eigenvalues $\tilde{\lambda}_1 \le \tilde{\lambda}_2 \le \ldots \le \tilde{\lambda}_m$ of $F_E$, and $|\lambda_i - \tilde{\lambda}_i| \le 2(\gamma /\veps)^2|\lambda_i - E|$.
}
\end{enumerate}
\label{fundamental}
\end{lemma}
Note that the degree of closeness between the two sets of eigenvalues improves the closer one gets to $\lambda = E$.

Our goal, then, is to iterate the fundamental lemma, using appropriately chosen subspaces, so that the spectral window width $\veps$ tends to zero, with $\gamma/\veps$ tending to zero as well. Eventually, the width will be so small that at most one eigenvalue will be inside. At that point, the decay of the eigenfunction will become manifest, as will the separation of the eigenvalue from the rest of the spectrum.

\tc{black}{
 The operator $- (D-\lambda)^{-1} C$ plays the role of expanding a solution of a low dimensional space to a higher dimensional
 space.} \tc{black}{ Such factors will be composed with each other to obtain the proper eigenfunction on the full Hilbert space; hence
  we call them {\it eigenfunction-generating kernels}.}

\textit{Proof.} The first statement is a standard fact about the Schur complement. If 
$\varphi$ is an eigenvector for $F_\lambda$ with eigenvalue $\lambda$, then
\begin{align}
(K-\lambda)\left( \begin{array}{cc}
\varphi \\- (D-\lambda)^{-1}C\varphi  \end{array} \right) &=\left( \begin{array}{cc}
(A-\lambda)\varphi - B(D-\lambda)^{-1}C\varphi  \\C\varphi - (D-\lambda)(D-\lambda)^{-1}C\varphi  \end{array} \right) \nonumber
\\& = 
\left( \begin{array}{cc}
(F_\lambda-\lambda)\varphi \\ 0  \end{array} \right) = 0.
\label{(1.7)}
\end{align}
Conversely, if $(K-\lambda)(\varphi,\tilde{\varphi}) = 0$, then $C\varphi + (D-\lambda)\tilde{\varphi} = 0$, which has unique solution $\tilde{\varphi} = -(D-\lambda)^{-1}C\varphi$. Substituting in, we obtain
$(F_\lambda - \lambda)\varphi = 0$.  Note that this implies agreement between the multiplicities of the $\lambda$-eigenspaces for $H$ and $F_\lambda$.

For the second statement, write
\begin{equation}
F_E-F_\lambda = -B(D-E)^{-1}C + B(D-\lambda)^{-1}C = B(D-E)^{-1}(\lambda-E)(D-\lambda)^{-1}C.
\label{(1.8)}
\end{equation}
Noting that $|\lambda-E| \le \veps/2$ and $\mathrm{dist(spec\,}D,E) \ge \veps$, we have that $\|(D-\lambda)^{-1}\| \le 2/\veps$, and thus
\begin{equation}
\|F_E-F_\lambda\| \le \frac{2\gamma^2}{\veps^2}|\lambda - E|.
\label{(1.9)}
\end{equation}
By Weyl's inequality, the eigenvalues of $F_\lambda$ and $F_E$ differ by no more than $2(\gamma /\veps)^2|\lambda_i - E|$ when shifting from $F_\lambda$ to $F_E = F_\lambda + (F_E-F_\lambda)$.\qed
\section{Fixed Energy Procedure}\label{sec:fixed}
\subsection{First Step}\label{ssec:first}

Let us investigate the spectrum of $H$ in the vicinity of $E$. The energy $E$ can be any real number (although since $H$ is bounded, there will be no spectrum near $E$ if $|E|$ is large). For the time being, $E$ will be treated as a fixed parameter.
\tc{black}{We emphasize that we are deriving probabilistic bounds on resonances by way of the geometric/algebraic behavior of  the 
eigenvalues with respect to the random parameters. Let us observe this in the simplest possible case.}

It will be helpful to introduce some terminology. The rectangle $\Lambda$ is a collection of \textit{positions} in $\mathbb{Z}^d$ labeling the coordinates in the first factor in $\calH=\mathbb{C}^{|\Lambda|}\otimes\mathbb{C}^2$. Coordinates for the full $\calH$ will be called \textit{sites}; thus there are $|\Lambda|$ positions and $2|\Lambda|$ sites. Let $\Lambda^*$ denote the set of sites. 
More generally, let $X^*$ denote the set of sites at positions $x\in X \subset \Lambda$.
It can be visualized as two copies of $\Lambda$ in two layers: two sites at each position. 

The eigenvalues of $h_x$ are $\pm t_x$, where
\begin{equation}
t_x = \sqrt{u_x^2+1}.
\label{(1.10)}
\end{equation}
We say that $x$ is $\veps$-resonant with $E$ if either eigenvalue is within $\veps$ of $E$. Sites associated with non-resonant positions are also said to be non-resonant.  The eigenvalues are split by at least 1, hence for $\veps$ small only one can be close to $E$. A direct calculation shows that
\begin{equation} 
P(t_x \in [|E|-\veps,|E|+\veps]) \le P(t_x \in [1,1+2\veps]) = 2\sqrt{\veps + \veps^2} \le 3\sqrt{\veps}, \text{  for  }\veps \le 1,
\label{(1.11)}
\end{equation}
assuming a uniform distribution on [-1,1] for $u_x$. (The maximum is achieved when $|E| = 1+\veps$.)
 Thus $3\sqrt{\veps}$ is a bound for the probability that $x$ is an $\veps$-resonant 
 position with respect to $E$. We take $\veps = \gamma^\phi$ with $\phi = \tfrac{1}{4}$.

Let $R^{(1)}$ denote the set of $\veps$-resonant positions $x \in \Lambda$. It can be broken into components, where connectedness is based on nearest-neighbor connections. 
  Let us estimate $\calP_{x,y}^{(1)}$, the probability that $x$ and $y$ are in the same  component.
   For positions $x$ and $y$ to be connected, there must be a self-avoiding walk from $x$ to $y$, consisting of nearest-neighbor steps between resonant sites.
  By (\ref{(1.11)}), the probability that all the sites in a given walk are resonant is bounded by $(3\sqrt{\veps})^m$, 
   where $m$ is the number of positions in the walk, including endpoints. If we add a combinatoric factor $2d$ per step, 
   we may convert the sum over walks with $m$ positions into a supremum\footnote{Combinatoric factors are a convenient bookkeeping device when estimating sums. If $\sum_\alpha c_\alpha^{-1} \le 1$ for some positive constants $c_\alpha$, then $\sum_\alpha X_\alpha \le \sup_\alpha |X_\alpha|c_\alpha$. We call $c_\alpha$ a combinatoric factor.}. 
  A combinatoric factor $2^m$ allows us to fix $m$. We obtain
\begin{equation}
\calP_{x,y}^{(1)} \le (12d\sqrt{\veps})^{m} \le (12d\sqrt{\veps})^{|x-y|+1},
\label{(1.12)}
\end{equation}
as $m-1$ must be at least $|x-y|$, the minimum number of steps to get from $x$ to $y$.

\textit{Block form of the Hamiltonian.} 
Recall that $R^{(1)}$ is the set of resonant positions. Putting $R^{(1)\text{c}} = \Lambda \setminus R^{(1)}$, we have associated sets of sites $R^{(1)*}$, $R^{(1)\text{c}*}$.
These index sets determine the block form of the Hamiltonian:
\begin{equation}
H = \left( \begin{array}{cc}
A^{(1)} & B^{(1)}\\  C^{(1)} & D^{(1)} \end{array} \right),
\label{(1.14)}
\end{equation}
with $A^{(1)}$ representing the restriction of $H$ to $R^{(1)*}$, and $D^{(1)}$ 
   representing the restriction to $R^{(1)\text{c}*}$ (or, to be precise, 
these are the restrictions of $H$ to the subspaces corresponding to those index sets). 
Note that $A^{(1)}, D^{(1)}$ contain off-diagonals only between sites 
at the same position or at
adjacent positions;
 positions which belong to different components in $R^{(1)}$ do not interact.

  We may write
\begin{equation}
D^{(1)} = W^{(1)} + V^{(1)},
\label{(1.15)}
\end{equation}
with $W^{(1)}$ block diagonal (with blocks $h_x$ at each position $x$) and with $V^{(1)}$ equal to the projection of $V$ to $R^{(1)\text{c}}$. 
 \tc{black}{Let $V^{(1)}_{xy}$ denote the  block matrix element of $V^{(1)}$  between positions $x$ and $y$. 
For $x, y \in R^{(1)\text{c}}$, it is equal to $\gamma J_{xy}I_2$, see (\ref{(1.3)}). Thus}
 it is nonzero only for \tc{black}{adjacent positions $x$ and $y$ in $R^{(1)\text{c}}$},
 \tc{black}{in which case it equals $\gamma I_2$.}

\tc{black}{
\textit{Convention.} We use ordinary absolute value signs $|\cdot|$ to denote the norm of a $2\times 2$ matrix such as the block matrix element $V^{(1)}_{xy}$. This is to distinguish it from the usual notation $\|\cdot\|$ for matrices indexed by $\Lambda^*$ or some subset of $\Lambda^*$. Using this notation, we have that $|V_{xy}^{(1)}|=\gamma$ for $x,y$ adjacent positions in $R^{(1)\text{c}}$.}

\textit{Random walk expansion}. 
\tc{black}{
We would like to reduce (\ref{(1.14)}) by  Schur complementation:
 \begin{equation}
F_{\lambda}^{(1)} \equiv A^{(1)} - B^{(1)}(D^{(1)}-\lambda)^{-1}C^{(1)}.
\label{(1.19)}
\end{equation}
\tc{black}{Let us develop a random walk expansion for} $ (D^{(1)}-\lambda)^{-1} $.
}

By construction, $\mathrm{dist(spec\,}\tc{black}{W}^{(1)},E) \ge \veps$. \tc{black}{Let us assume that $|\lambda-E| \le \veps/2$,  so that  $\|(W^{(1)}-\lambda)^{-1}\|\le 2/\veps$. }
 The off-diagonal matrix $V^{(1)}$ has norm no greater than $\tc{black}{2}d\gamma$ in dimension $d$ (it is a symmetric matrix with row sum equal to $\tc{black}{2}d\gamma$). 
Therefore, \tc{black}{ for $\veps = \gamma^\phi = \gamma^{1/4}$ sufficiently small}, $(D^{(1)} - \lambda)^{-1}$
  exists and can be expanded in a Neumann series: 
\begin{equation}
(D^{(1)} - \lambda)^{-1} = (W^{(1)} - \lambda)^{-1} + (W^{(1)} - \lambda)^{-1}V^{(1)}(W^{(1)} - \lambda)^{-1} + \ldots.
\label{(1.16)}
\end{equation}
This leads to a random walk expansion with steps determined by \tc{black}{block matrix elements of} $V^{(1)}$: 
\begin{equation}
\left[(D^{(1)} - \lambda)^{-1}\right]_{xy} = \sum_{g_1:x\rightarrow y} \prod_{i=1}^m [(W^{(1)} - \lambda)^{-1} ]_{x_ix_i} \prod_{j=1}^{m-1} V^{(1)}_{x_jx_{j+1}}.
\label{(1.17)}
\end{equation}
\tc{black}{Here $g_1 = \{x=x_1, x_2, \ldots, {x}_m = y\}$ is a random walk with each $x_i$, $x_{i+1}$  adjacent positions; return visits are allowed. }
 \tc{black}{For such a walk we write its length as $|g_1| = m$.}
\tc{black}{Note that each factor on the right-hand side of (\ref{(1.17)}) is a $2\times2$ block matrix element.}
As each eigenvalue of $W^{(1)}-\lambda$ is at least $\veps/2$ in magnitude, we have that 
 $|[(W^{(1)} - \lambda)^{-1} ]_{x_ix_i} | \le 2/\veps$ for every $x_i \in g_1$.
  Thus a term in the sum \tc{black}{(\ref{(1.17)})} with $|g_1|=m$ is bounded by $\gamma^m(\veps/2)^{-(m+1)}$.
   \tc{black}{We may assign combinatoric factors $2^m$ to fix the length of the walk 
    and $2d$ for each sum over $x_j$};
 \tc{black}{altogether the combinatoric factor is $2^m(2d)^{m-1}$.}
 \tc{black}{Thus}  we obtain a bound
\begin{equation}
\left| \left[(D^{(1)} - \lambda)^{-1}\right]_{xy}\right| \le \left(\frac{c_d\gamma}{\veps}\right)^{|x-y|}\cdot\frac{4}{\veps} ,
\label{(1.18)}
\end{equation}
as $m-1$ must be at least $|x-y|$, the number of steps needed to walk from $x$ to $y$. Here, and in what follows, $c_d$ denotes a constant that depends only on the dimension $d$.

 \tc{black}{Observe from (\ref{(1.16)}) and (\ref{(1.17)}) that the operator $(D^{(1)} - \lambda)^{-1}$ connects all sites in  $R^{(1)\text{c}*}$. 
 Therefore, we see that resonant sites $R^{(1)*}$ are connected through the chain of operators $B^{(1)}(D^{(1)}-\lambda)^{-1}C^{(1)}$ in the second term of (\ref{(1.19)}). Note that
 $B^{(1)}$ and $C^{(1)}$ contain  off-diagonal interactions from $V$.
  Thus the walks in (\ref{(1.17)}) are extended with two additional steps $V^{(1)}_{x_jx_{j+1}}$. As in the arguments for (\ref{(1.18)}), we obtain}
 \begin{equation}
\left|\left[B^{(1)}(D^{(1)}-\lambda)^{-1}C^{(1)}\right]_{xy}\right| \le \left(\frac{c_d\gamma}{\veps}\right)^{|x-y|\vee2}\cdot\veps.
\label{(1.20)}
\end{equation}
In a similar manner, we can estimate the \tc{black}{\it eigenfunction-generating kernel} \tc{black}{for the first step}:
\begin{equation}
\left|\left[(D^{(1)}-\lambda)^{-1}C^{(1)}\right]_{xy}\right| \le \left(\frac{c_d\gamma}{\veps}\right)^{|x-y|\vee1}.
\label{(1.21)}
\end{equation}
The graphical expansion (\ref{(1.17)}) induces similar expansions for these matrix elements. 

\subsection{Resonant Blocks}\label{ssec:resonant}
   In the $k^{\text{th}}$ step we use a length scale $L_k \equiv 2^{k-1}$ to determine connectivity. 
   \tc{black}{That is,} in the first step, $L_1 = 1$ and two 
    \tc{black}{positions are in the same component} if they are nearest neighbors. 
   In steps $k \ge 2$, we inherit a set of resonant positions $R^{(k-1)}$,
  \tc{black}{and two positions are in the same component if they are within a distance $L_k$.
   Thus, components at scale $ L_k$ are subsets of $R^{(k-1)}$ that can be spanned by connections of length up to $L_k$.}
  A component of \tc{black}{cardinality} $n$ has $2n$ sites; each of these is effectively connected to all the others of the component. 
    Distance will always be measured in terms of the original metric $|x| = \sum_{i=1}^d |x_i|$ on $\mathbb{Z}^d$. 

\begin{definition}
Let $B$ be a component of $R^{(k-1)}$ on scale $k$ with $k\ge2$. We say that $B$ is  \textbf{isolated on scale k}
  if
\begin{enumerate}
\item{its \textbf{volume} ($\equiv n(B) \equiv$ number of positions) is no greater than $L_k^{2\psi/5}$,}
\item{its distance from every other component on scale $k$ is at least $4L_k$,}
\item{its diameter is $\le L_k$.}
\end{enumerate}
\label{def:isolated}
Here $\psi = \tfrac{2}{3}$ is an exponent governing fractional exponential decay of \tc{black}{graphical bounds}.
\end{definition}
From the previous step, we have the Schur complement matrix
\begin{equation}
F_{\lambda}^{(k-1)} = A^{(k-1)} - B^{(k-1)}(D^{(k-1)}-\lambda)^{-1}C^{(k-1)},
\label{(1.22)}
\end{equation}
with bounds similar to (\ref{(1.20)}) that are proven below in Theorem \ref{thm:2} (and that can be assumed by induction). It operates on the space indexed by elements of $R^{(k-1)*}$. Each of the matrices in (\ref{(1.22)}) is given by a random walk expansion as in (\ref{(1.17)}) in terms of the corresponding matrices from step $k-2$. 
See Section \ref{ssec:rwe} for details on the random walk expansion; for now, we need only some of its general properties. 

\tc{black}{In each step we take the Schur complement of the current renormalized Hamiltonian. 
               In the first step, $F^{(1)}$ is a renormalization of the original Hamiltonian $H$;
               the interactions of the first step generate walks, or {\it graphs} constructed of connections in the original Hamiltonian.
               Further renormalizations $F^{(2)},F^{(3)},\ldots$ involve walks of the preceding Hamiltonian's walks; we will 
              refer to these nested walk structures as {\it multigraphs}.}
Thus the  Schur complement $F_\lambda^{(k-1)}$ is defined inductively in terms of $(k-1)$-level \textit{multigraphs}, 
 each being a walk with steps given by $(k-2)$-level multigraphs, and so on down to the first level. By construction, these graphs sample the potential $u_x$ only for 
\tc{black}{ positions $x$ that the first-level walk passes through; thus there is dependence on $u_x$ only for}
 $x \notin R^{(k-1)}$. Consequently, the dependence on $u_x$, $x \in R^{(k-1)}$ is only through the leading term in $F_\lambda^{(k-1)}$, specifically the matrix $A^{(k-1)}$, projected to $R^{(k-1)*}$. (Recall that $A^{(1)}$ is the original Hamiltonian $H$ projected to $R^{(1)*}$.)
 \tc{black}{In particular, the values $u_x \in R^{(k-1) } $ {\it do not} appear in any }\tc{black}{of the block inverse operators used -- see (\ref{(1.17)}) and its $k^{\text{th}}$ step version (\ref{(1.50)}).} 

 One may visualize the
last term in (\ref{(1.22)}) as a set of graphs that exit $R^{(k-1)}$ (the matrix $B^{(k-1)}$), and then wander in $R^{(k-2)}\setminus R^{(k-1)}$ (the matrix
$(D^{(k-1)}-\lambda)^{-1}$) before returning to $R^{(k-1)}$ (the matrix $C^{(k-1)}$).
Thus we see that $F_\lambda^{(k-1)}$ agrees with $H$ on $R^{(k-1)*}$ up to small terms from the random walk expansions, all independent of $u_x$, $x \in R^{(k-1)}$. 
   Hence $F_\lambda^{(k-1)}$  \tc{black}{is \tc{black}{equal} to a direct sum of $H$ projected to the $R^{(k-1)}$  components} 
   plus small $O(\gamma^2/\veps)$ corrections (independent of $u_x$, $x \in R^{(k-1)}$).
  \tc{black}{ That is, we may write} 
  \begin{equation}
   \label{fk general form}  \tc{black}{
    F_\lambda^{(k-1)} = \oplus_B h_B + \Xi_{k-1},  }
\end{equation}
\tc{black}{  where $h_B$ is the projection of $H$ to the component  $B$ in $R^{(k-1)}$ 
   and $\tc{black}{\Xi_{k-1}} = O(\gamma^2/\veps)$ is  independent of $u_x$, $x \in R^{(k-1)}$.
  For $k = 2$ (\ref{fk general form}) holds by (\ref{(1.20)}) 
   and indeed we will generalize it for $k > 2$ in the induction procedure -- see }\tc{black}{ Corollary \ref{cor:2'}.}
Furthermore, there is no off-diagonal dependence on $u_x$, $x \in R^{(k-1)}$.

We wish to show that  isolated components are unlikely to contain spectrum near $E$. 
We could work with $F_\lambda^{(k-1)}$, projected to an   isolated  component $B$ of $R^{(k-1)}$, but in order to preserve independence we truncate the random walk expansions in  $F_\lambda^{(k-1)}$, including only \tc{black}{multi}graphs that remain within a distance $< L_{k-2}$ of $B$. \tc{black}{Multi}graphs neglected therefore travel a distance $\ge 2L_{k-2} = L_{k-1}$ after returning to $B$ or reaching another block.
  \tc{black}{Let us designate such multigraphs {\it nonlocal} at scale $L_{k-1}$; they visit points at a distance $\ge L_{k-2}$ from $B$.} \tc{black}{
Let us write $\tilde F_E^{(k-1)}$ for the renormalized Hamiltonian with all nonlocal multigraphs dropped.
 The truncated Hamiltonian is block diagonal due to the separation condition: $\tilde F_E^{(k-1)} = \oplus_B \tilde F_E^{(k-1)}(B) $.
\tc{black}{As above, we may} write  $\tilde{F}_\lambda^{(k-1)}(B) =  h_B + \tilde \Xi_{k-1} (B)  $
 for the truncated $F_\lambda^{(k-1)}$, projected to block $B$, \tc{black}{and $\tilde \Xi_{k-1} (B) = O(\gamma^2/\veps)$}}.
 
We prove decay at rate $\gamma^{L^\psi}$ 
for graphs of length $L$, with $\psi = \tfrac{2}{3}$ -- see Theorem \ref{thm:2} below. 
We \tc{black}{prove below} a bound on the probability of spectrum within $\veps^{L_{k-1}^\psi}$ of $E$. As this window is much larger than the size of terms neglected (recall that $\veps = \gamma^\phi$ with $\phi = \tfrac{1}{4}$), movement of the spectrum due to terms of order $\gamma^{L_{k-1}^\psi}$ is insignificant. 

\begin{definition}
Let $B$ be a component of $R^{(k-1)}$ on scale $k$ with $k\ge2$. We say that $B$ is \textbf{resonant on scale k} 
  if it is isolated on scale $k$ and if $\mathrm{dist(spec\,}\tilde{F}_E^{(k-1)}(B),E) \le \veps_k \equiv \veps^{L_k^\psi}$.
 Here we refer to the truncation $F^{(k-1)}_E \rightarrow \oplus_B\tilde{F}^{(k-1)}_E(B)$, which results from the deletion of all multigraphs that extend a distance $L_{k-2}$ or farther from any block $B$. 
\label{def:resonant}
\end{definition}
Following a similar plan as in the first step,
 we define the new resonant set $R^{(k)}$ by removing from $R^{(k-1)}$ all the non-resonant isolated components.
 Thus
\begin{equation}
R^{(k)}=R^{(k-1)} \setminus  \bigcup_{\alpha:\,B_\alpha \text{ is  isolated  but not resonant on scale }k }B_\alpha.
\label{(1.28)}
\end{equation}
\begin{proposition}
Let $B$ be an isolated
    component of $R^{(k-1)}$ on scale $k$ with $k\ge2$. 
The probability that  $B$ is resonant on scale $k$ is less than $\veps^{L_k^{3\psi/5}/3}$.
\label{prop:A}
\end{proposition}

\textit{Proof}. Let us enumerate the positions in $B$ as $\{x_1,\ldots,x_n\}$. The block $B$ is connected on scale $L_{k-1}$ (\textit{i.e.} using connections between $x_i$ and $x_j$ of length up to $L_{k-1}$. 
Let us consider the $2n\times2n$ matrix $\tilde{F}_\lambda^{(k-1)}(B)$ as a function of $\mathbf{u}=\{u_{x_1},\ldots,u_{x_n}\}$, 
with all other $u$'s held fixed, \tc{black}{which means that $\tilde \Xi_{k-1} (B)$ is fixed as well}. 
Thus, as explained above, $\tilde{F}_\lambda^{(k-1)}(B)$ agrees with $H$ up to $\mathbf{u}$-independent terms of size $O(\gamma^2/\veps)$. 
We consider here $\lambda = E$, since by the fundamental lemma, control of the spectrum of $\tilde{F}_E^{(k-1)}(B)$ allows us to control $\tilde{F}_\lambda^{(k-1)}(B)$ for $\lambda$ near $E$. Recall that $H$ has $u_i$ or $-u_i$ on the diagonal, and it has off-diagonal matrix elements connecting adjacent positions and connecting the two sites at each position. The off-diagonal matrix entries are either $\gamma$, 1, or 0. It should be clear, then, that $\Delta(\mathbf{u}) \equiv \text{det}(\tilde{F}_E^{(k-1)}(B)-E)$ is a polynomial in $\mathbf{u}$ of degree $2n$. We may therefore apply standard results bounding the measure of sets where the polynomial is small.

Noting that each $u_x \in [-1,1]$, and since $t_x = \sqrt{u_x^2+1}$, the spectrum of $H$ is clearly limited to the range $[-\sqrt{2},\sqrt{2}]+O(\gamma)$. Therefore, we can assume $E \in [-\tfrac{3}{2},\tfrac{3}{2}]$; this will cover all of the spectrum of $H$. 

\tc{black}{With $E \in [-\tfrac{3}{2},\tfrac{3}{2}]$, we need to establish in a quantitative sense that $\Delta(\mathbf{u})=\text{det}(\tilde{F}_E^{(k-1)}(B)-E)$ is not the zero polynomial. Thus we look for a value of}
 $\mathbf{u}$ for which $\Delta(\mathbf{u})$ is bounded away from 0. Take $\mathbf{u}_0  = (2,\ldots,2)$. Then each $t_{x_i}=\sqrt{5}$, and so
\begin{equation}
(t_x-E)(-t_x-E) = E^2-5 \le \tfrac{9}{4}  - 5 = -\tfrac{11}{4}.
\label{(1.23)}
\end{equation}
We must allow for $O(\gamma)$ movement of the eigenvalues due to the terms other than $\oplus_i h_{x_i}$ in $\tilde{F}_E^{(k-1)}(B)$ (here we use also the decay away from the diagonal -- see Theorem \ref{thm:2} below -- and the fact that the norm of the perturbing matrix is bounded by the maximum absolute row sum). Still, we maintain a lower bound $|\Delta(\mathbf{u}_0)| \ge 2^n$, after taking the product over the $n$ $2\times2$ blocks and allowing for the small changes in the eigenvalues.

We use the Brudnyi-Ganzburg inequality, in the form stated in \cite{Brudnyi1999}:
\begin{equation}
\underset{U}{\text{sup}} \,|p| \le \left(\frac{4n|U|}{|\omega|}\right)^\kappa \underset{\omega}{\text{sup}} \,|p|.
\label{(1.24)}
\end{equation}
Here $U$ is a bounded convex subset of $\mathbb{R}^n$, $p$ is a polynomial of degree at most $\kappa$ and $\omega$ is a measurable subset of $U$. In our case, we let $U = [-2,2]^n, \kappa = 2n$, and
\begin{equation}
\omega = \{\mathbf{u}:\, |\Delta(\mathbf{u})| \le \veps_k 3^{2n} \} \text{ with } \veps_k\equiv \veps^{L_k^\psi}.
\label{(1.25)}
\end{equation}
We have shown that $\text{sup}_U\,|\Delta(\mathbf{u})| \ge 2^n$ and so
\begin{equation}
|\omega| \le \left(2^{-n}\veps_k 3^{2n}\right)^{\tfrac{1}{2n}}\cdot 4n|U| \le \veps_k^{1/(2n)} 6\sqrt{2}\cdot n4^n.
\label{(1.26)}
\end{equation}
Since $u_x$ is distributed uniformly on [-1,1], this leads to bounds
\begin{align}
P(|\Delta(\mathbf{u})| \le \veps_k 3^{2n} ) &\le \veps^{L_k^\psi/(2n)} \cdot 9n2^n \nonumber
\\&\le\veps^{L_k^{3\psi/5}/2}\cdot 9L_k^{2\psi/5}2^{L_k^{2\psi/5}} \le \veps^{L_k^{3\psi/5}/3},
\label{(1.27)}
\end{align}
where in the second line we have used the bound $n \le L_k^{2\psi/5}$ from Definition \ref{def:isolated} (as well as the smallness of $\veps = \gamma^\phi$).
Note that both $E$ and the eigenvalues of $\tilde{F}_E^{(k-1)}(B)$ are in $[-\tfrac{3}{2},\tfrac{3}{2}]$, and therefore the eigenvalues of $\tilde{F}_E^{(k-1)}(B)-E$ are no greater than 3 in magnitude. Hence (\ref{(1.27)}) implies the statement of the proposition.\qed

\subsection{Probability Estimates}\label{ssec:prob}
We need to work on a multiscale version of the percolation estimate (\ref{(1.12)}). As we investigate the percolation of resonant blocks, we face a situation in which blocks of $R^{(k)}$ may inherit probability bounds from step $k - 1$ or they may receive a probability bound from Proposition \ref{prop:A}. Thus we make an inductive definition that keeps track of the probability bounds available for blocks at any scale:
\begin{align}
\tilde{P}^{(1)}(B) &\equiv (3\sqrt{\veps})^{n(B)} \nonumber
\\
\tilde{P}^{(k)}(B) &\equiv \begin{cases}\veps^{L_{k-2}^{3\psi/5}/3}, & \text{if }B\text{ is isolated on scale }k - 2;\\
\prod_{i=1}^{m'} \tilde{P}^{(k-1)}(B_i), &\text{otherwise}.
\end{cases}
\label{(1.30)}
\end{align}
\tc{black}{We emphasize that the $\tilde P$ are not themselves probabilities but stand in for upper bounds on the probabilities of resonant clusters.}
Here $n(B)$ is the number of positions in $B$, and $B_1,\ldots,B_{m'}$ are the subcomponents of $B$ on scale $k-1$. Note that each factor $3\sqrt{\veps}$ or  $\veps^{L_{k-2}^{3\psi/5}/3}$ (corresponding to a bound on the probability that a site or block is resonant) carries forward to the next scale, until a scale $k$ is reached
with $B$ isolated on scale $k-2$. In that case, all the factors from subblocks
of $B$ from previous scales are replaced with the bound from
Proposition \ref{prop:A}. 
\tc{black}{This corresponds to replacement of bounds by a better one from Proposition \ref{prop:A}, when it is available (\textit{i.e.} if a component is isolated on scale $k-2$).}

Next let us define a weighted sum of the probability bounds $\tilde{P}^{(k)}(B)$:
\begin{equation}
Q^{(k)}_{x,y} \equiv \sum_{B\text{ containing }x\text{ and }y}  \tilde{P}^{(k)}(B)\veps^{-q_k n(B)}.
\label{(1.31)}
\end{equation}
Here $\{q_k\}$ is a decreasing sequence of numbers with a strictly positive floor $q>0$; it will be specified in the next theorem.  We prove the following 
``energy-entropy'' bound, similar in spirit to \cite{Frohlich1983}. It is a purely 
geometric-combinatoric result about the  components $B$ that are generated by our definitions.

\begin{theorem}
Given $\psi=\tfrac{2}{3}$, let $\chi = \psi/3$. Then for $\veps$ sufficiently small, and $q_1 \equiv \tfrac{1}{5}$, $q_2 \equiv \tfrac{1}{6}$, $q_3 \equiv \tfrac{1}{7}$, and $q_k \equiv q_{k-1}(1-L_{k-2}^{-\psi/20})$ for $k \ge 4$, we have
\begin{equation}
Q^{(k)}_{x,y} \le \veps^{q_k (|x-y|\vee L_{k-2})^\chi}.
\label{(1.32)}
\end{equation}
\label{thm:1}
\end{theorem}

\textit{Proof.} To treat the case $k=1$, we need to extend the proof of (\ref{(1.12)}) to incorporate decay with $n=n(B)$, the volume of the component containing $x,y$. The walk from $x$ to $y$ can be extended to a branching random walk (\textit{i.e.} a spanning tree). The number of such trees is bounded by $c_d^n$. The weighting $\veps^{-q_1 n} = \veps^{-n/5}$ in (\ref{(1.31)}) is counterbalanced by our bound on the probability that $n$ positions are resonant, $(3\sqrt{\veps})^n$. Noting that $n \ge |x-y|+1$, we may extract the decay in (\ref{(1.32)}), with a decrease of an additional 1/5 in the power of $\veps$. This leaves a bound of $(2c_d\veps^{1/10})^n$. The sum on $n$ is less than 1, and hence (\ref{(1.32)}) holds for $k=1$.

\begin{figure}[h]
\centering
\includegraphics[width=.8\textwidth]{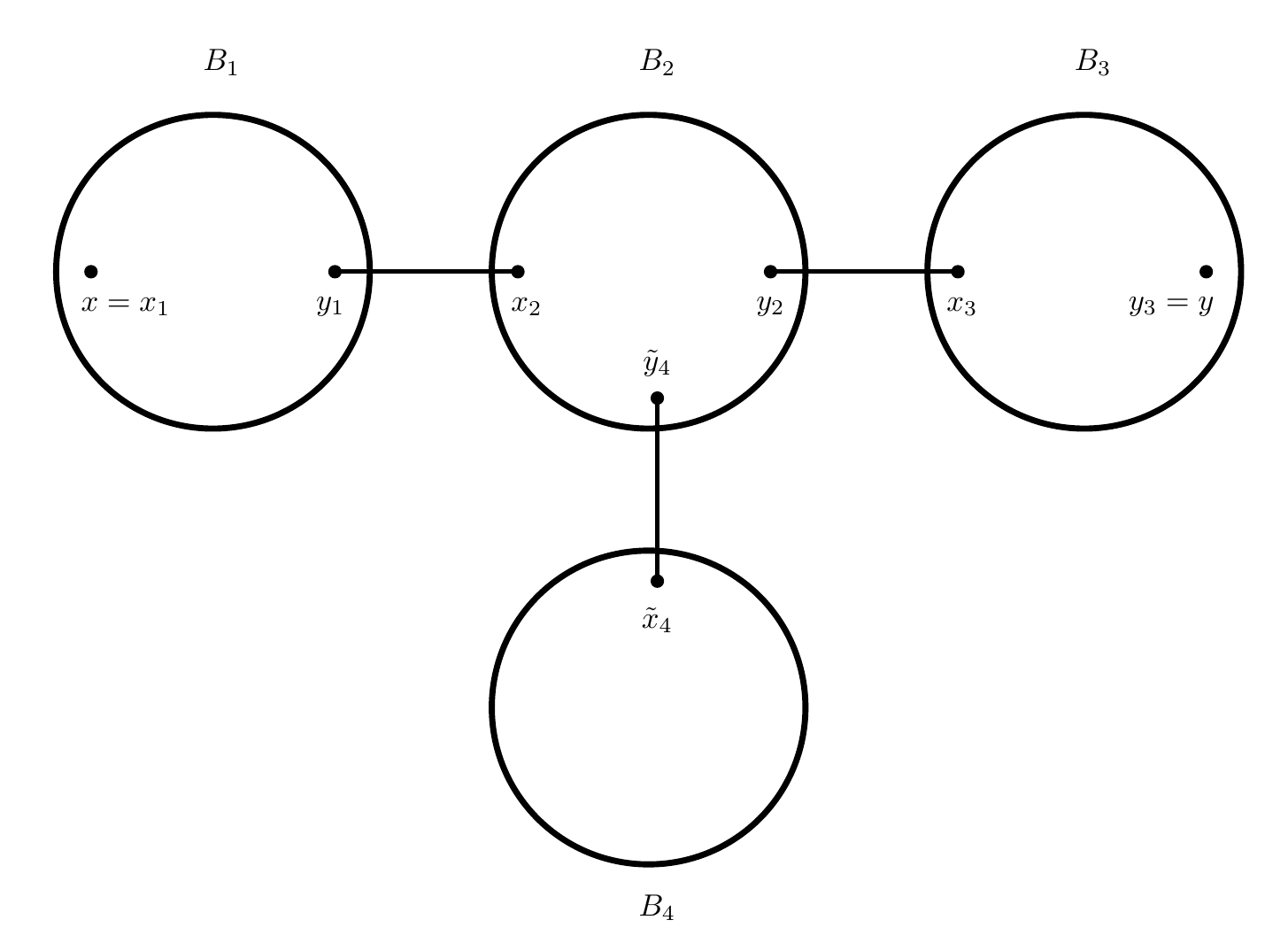}
\caption{A tree graph of blocks connecting $x$ and $y$. Blocks $B_1, B_2, B_3$ form the spine and $B_4$ is a leaf.} \label{fig:1}
\end{figure}

Let us now assume (\ref{(1.32)}) for $k-1$. Let us write
\begin{equation}
Q^{(k)}_{x,y} = Q^{(k),1}_{x,y} + Q^{(k),2}_{x,y},
\label{(2.21a)}
\end{equation}
with the first term giving the sum over blocks $B$ that are \textit{not}  isolated on 
scale $k-2$, and the second term giving the sum over the ones that \textit{are} isolated. We consider first
$Q^{(k),1}_{x,y}$, and note that $\tilde{P}^{(k)}(B)$ is given by the
second \tc{black}{alternative} in (\ref{(1.30)}). Recall that $B$ is a component of $R^{(k)}$ and that it is formed by joining together components of $R^{(k-1)}$ with links of length
$\le L_k$. Hence there
 must be a sequence of positions $x=x_1,y_1,x_2,y_2,\ldots,x_m,y_m = y$ such that each $x_i,y_i$ lie in the same component $B_i$ of $R^{(k-1)}$, and $|x_i-y_{i-1}| \le L_k$. The remaining subcomponents of $B$ may be denoted $B_{m+1},\ldots,B_{m'}$. Connectivity on scale $k$ implies that there must be a tree graph $T$ connecting all the components of $B$. Each link of the tree graph is a pair $(x_i, y_{i-1})$ as above with $2 \le i \le m$ or a pair $(\tilde{x}_j,\tilde{y}_j)$ with $m+1 \le j \le m'$, $\tilde{x}_j \in B_j$, $\tilde{y}_j \in B_{\tau(j)}$, and $\tau(j) < j$. Thus we choose a tree graph that extends the unbranched connection from $B_1$ to $B_m$. (See Figure \ref{fig:1}.)
Again, because of connectivity, we can assume that $|\tilde{x}_j - \tilde{y}_j| \le L_k$.

We may write
\begin{equation}
Q^{(k),1}_{x,y} \le \sum_{m,m'} \sum_T\, \sum_{\substack{B_1,\ldots,B_{m'}\\ \text{consistent with }T }}\,
\prod_{i=1}^{m'}\left[ \tilde{P}^{(k-1)}(B_i) \veps^{-q_k n(B_i)}\right],
\label{(2.22)}
\end{equation}
where we have written $n=\sum_i n(B_i)$ as the sum of the volumes of the $B_i$. With some overcounting, we may proceed through the tree graph, summing over vertices $x_i,y_i,\tilde{x}_j,\tilde{y}_j$ and components, requiring only that each $B_i$ contain a specified set of 1 or 2 vertices. In this way, we will be able to apply the inductive
hypothesis from step $k-1$.

The tree graph $T$ may be broken up into the chain connecting $B_1,\ldots,B_m$
plus individual tree graphs $T_1,\ldots,T_m$ with roots at the corresponding
blocks. We now focus on a single $B_i$ (which we take as fixed) and sum over
$T_i$ and the associated blocks (other than $B_i$) that are linked by $T_i$.
Let us define for an arbitrary block $\bar{B}$ on scale $k-1$:
\begin{equation}
K^{(p)}(\bar{B}) \equiv \sum_{\bar{T}: \text{ depth}(\bar{T})\le p}\,  \sum_{\substack{\bar{B}_1,\ldots,\bar{B}_{\ell}\\ \text{consistent with }\bar{T} }}\,
\prod_{i=1}^{\ell}\left[ \tilde{P}^{(k-1)}(\bar{B}_i) \veps^{-q_k n(\bar{B}_i)}\right].
\label{(2.22a)}
\end{equation}
Here $\text{depth}(\bar{T})$ is the largest number of links in $\bar{T}$ that are needed to reach a
vertex, starting at the root. 
Then we can write
\begin{align}
Q^{(k),1}_{x,y} &\le \sum_{m=1}^{\infty}\,\sum_{y_1,x_2,y_2,\ldots,y_{m-1},x_m} \,\prod_{i=1}^m 
\left[
\sum_{B_i\text{ containing }x_i,y_i}  \tilde{P}^{(k-1)}(B_i)\veps^{-q_k n(B_i)} K^{(\infty)}(B_i)
\right].
\label{(2.22aa)}
\end{align}
Here the inequality results from the overcounting associated with treating each of the $T_i$ as 
independent of the others. 
Let us put
\begin{equation}
\hat{\veps} = \veps^{q_{k-1}L_{k-3}^\chi/2}, \,k \ge 2.
\label{(2.22b)}
\end{equation}
\begin{lemma}
\label{lem:polymer}
With $\chi = \tfrac{2}{9}$ and $\veps$ sufficiently small, 
\begin{equation}
K^{(p)}(\bar{B}) \le e^{\hat{\veps}n(\bar{B})}.
\label{(2.22c)}
\end{equation}
\end{lemma}
\textit{Proof}. We use induction on $p$. When $p=0$, we have the tree with no
links, so (\ref{(2.22c)}) amounts to the statement $1 \le e^{\hat{\veps}n(\bar{B})}$.
For larger $p$, we may use the recursion
\begin{equation}
K^{(p)}(\bar{B}) \le \sum_{r=0}^\infty\frac{1}{r!} \prod_{i=0}^{r} \Bigg[\sum_{\bar{B}_i \text{ linked to }\bar{B}}\,
\tilde{P}^{(k-1)}(\bar{B}_i) \veps^{-q_k n(\bar{B}_i)} K^{(p-1)}(\bar{B}_i)\Bigg].
\label{(2.22d)}
\end{equation}
Again, this is an upper bound, as we are treating each of the
$r$ branches of $\bar{T}$ as independent of the others. There is also overcounting 
due to permutations of the branches, and this has been compensated by the factor
$1/r!$. 

The sum over $\bar{B}_i$ linked to $\bar{B}$ may be bounded as follows. By connectivity, there must be a pair of points $\bar{x}$, $\bar{y}$ with $\bar{y} \in \bar{B}$
and $\bar{x} \in \bar{B}_i$ and $|\bar{x} - \bar{y}| \le L_k$. The sum over 
$\bar{y}$ leads to a factor $n(\bar{B})$. There are no more than $(2L_k+1)^d$
choices for $\bar{x}$, given $\bar{y}$. \tc{black}{Furthermore,} we may bound
\begin{equation}
\veps^{-q_k n(\bar{B}_i)} K^{(p-1)}(\bar{B}_i) \le \veps^{-q_{k-1} n(\bar{B}_i)} 
\label{(2.22e)}
\end{equation}
using (\ref{(2.22c)}), since
\begin{align}
-(q_k - q_{k-1})|\log\veps| + \hat{\veps} &\ge 
\begin{cases}
|\log\veps| q_{k-1}L_{k-2}^{-\psi/15}, &k\ge 4 \\0, &k=2,3
\end{cases}\nonumber\\
&\ge 0
\label{(2.22f)}
\end{align}
for $\veps$ small. (This follows from the expressions for $q_k$ in Theorem \ref{thm:1}, in particular the relation $q_{k-1}-q_k = q_{k-1}L_{k-2}^{-\psi/20}$ for $k \ge 4$; 
the increase from $\psi/20$ to $\psi/15$ is to accommodate the $\hat{\veps}$ term
on the left-hand side.) Inserting (\ref{(2.22e)}) into (\ref{(2.22d)}), we obtain
\begin{align}
K^{(p)}(\bar{B}) &\le \exp\left((2L_k+1)^d Q_{\bar{x},\bar{x}}^{(k-1)}n(\bar{B})\right) \nonumber
\\
&\le \exp\left((2L_k+1)^d  \veps^{q_{k-1}L_{k-3}^\chi}   n(\bar{B})\right) \le e^{\hat{\veps}n(\bar{B})},
\label{(2.22g)}
\end{align}
after applying the induction hypothesis (\ref{(1.32)}) to $Q_{\bar{x},\bar{x}}^{(k-1)}$.
This closes the induction and completes the proof of the lemma. 
This ``polymer expansion'' construction is a standard way to treat tree graph sums
when the vertices have spatial extent, see for example \cite{Glimm1987}.\qed

After using the lemma to sum over all the branches of the tree, we are left with the
``spine,'' that is, the blocks $B_1,\ldots,B_m$ that lead from $x$ to $y$. \tc{black}{From (\ref{(2.22aa)}),} we obtain the following bounds:
\begin{align}\label{(1.33)}
Q^{(k),1}_{x,y} &\le \sum_{m=1}^{\infty}\,\sum_{y_1,x_2,y_2,\ldots,y_{m-1},x_m} \,\prod_{i=1}^m 
\left[
\sum_{B_i\text{ containing }x_i,y_i}  \tilde{P}^{(k-1)}(B_i)\veps^{-q_k n(B_i)} e^{\hat \veps n(B_i)}
\right]    \\
 &\le \sum_{B_1\text{ containing }x\text{ and }y}  \tilde{P}^{(k-1)}(B_1)\veps^{-q_k n(B_1)} e^{\hat{\veps}n(B_1)}
 +\sum_{m \ge 2} \,\sum_{y_1,x_2,y_2,\ldots,y_{m-1},x_m} \,\prod_{i=1}^m Q_{x_i,y_i}^{(k-1)},  \nonumber
\end{align}
The effect of the tree graphs has been subsumed into factors $e^{\hat{\veps}n(B_i)}$, by Lemma \ref{lem:polymer}. 
The first term \tc{black}{of the final expression} bounds the case $m=1$. 
For the terms $m \ge 2$, we have used the fact that
\begin{equation}\label{(1.33a)}
\sum_{B_i\text{ containing }x_i,y_i}  \tilde{P}^{(k-1)}(B_i)\veps^{-q_k n(B_i)} e^{\hat \veps n(B_i)} \le Q_{x_i,y_i}^{(k-1)},
\end{equation}
\tc{black}{which follows from (\ref{(1.31)}),(\ref{(2.22f)}).}

Consider the term $m=2$, and let $\tilde{\veps} \equiv \veps^{q_{k-1}}$. We need to be careful with the sums on $y_1,x_2$ in order to preserve the form of the decay estimate (\ref{(1.32)}). With $x,y$ fixed, note that
\begin{equation}
\sum_{y_1,x_2} \min\{|x-y_1|,|x_2-y|\}^{-(d+1)} \le c_d L_k^d.
\label{(1.34)}
\end{equation}
This can be seen by considering two cases, depending on whether the minimum is $|x-y_1|$ or $|x_2 - y|$, and noting that $|y_1-x_2| \le L_k$. As a consequence, a combinatoric factor
\begin{equation}
c_dL_k^d\min\{|x-y_1|,|x_2-y|\}^{(d+1)} \le \tilde{c}_d L_{\min}^{2d+1}
\label{(1.35)}
\end{equation}
suffices to control the sums. Here $L_{\min} \equiv \min\{L_a,L_b\}$, $L_{\max} \equiv \max\{L_a,L_b\}$, with $L_a \equiv |x-y_1|\vee L_{k-3}$, $L_b \equiv |x_2-y|\vee L_{k-3}$. Hence we may bound the $m=2$ term by
\begin{equation}
\sum_{y_1,x_2} Q_{x,y_1}^{(k-1)}Q_{x_2,y}^{(k-1)} \le \sup_{y_1,x_2} \tilde{c}_d L_{\min}^{2d+1} Q_{x,y_1}^{(k-1)}Q_{x_2,y}^{(k-1)} \le \sup_{y_1,x_2} \tilde{c}_d L_{\min}^{2d+1}\tilde{\veps}^{L_a^\chi}\tilde{\veps}^{L_b^\chi}.
\label{(1.36)}
\end{equation}

Note that
\begin{equation}
|x-y|\vee L_{k-2} \le L_a+L_k+L_b = L_{\min} + L_k + L_{\max},
\label{(1.37)}
\end{equation}
since as indicated above, $|x_2-y_1| \le L_k$ by connectivity of $B$. Let us write
\begin{equation}
\tilde{c}_d L_{\min}^{2d+1} \tilde{\veps}^{L_a^\chi}\tilde{\veps}^{L_b^\chi} = \tilde{c}_d L_{\min}^{2d+1} \tilde{\veps}^{L_{\min}^\chi}\tilde{\veps}^{L_{\max}^\chi} \le \tfrac{1}{4}\tilde{\veps}^{(4/5)L_{\min}^\chi}\tilde{\veps}^{L_{\max}^\chi},
\label{(1.37a)}
\end{equation}
which holds for small $\tilde{\veps}$. We claim that
\begin{equation}
\tfrac{4}{5}L_{\min}^\chi + L_{\max}^\chi > (L_{\min} + L_k + L_{\max})^\chi.
\label{(1.37b)}
\end{equation}
The worst case for this inequality is when $L_a,L_b$ are at their minimum possible values, $L_{k-3}$ (as one can check by differentiating). Then it reduces to the inequality $1.8 > 10^\chi$, or $\chi < \log_{10}1.8 \approx .255$. Recalling that $\psi = \tfrac{2}{3}$ and $\chi = \psi/3$, we can confirm this inequality, and then (\ref{(1.37)}),(\ref{(1.37b)}) imply a bound
\begin{equation}
\tfrac{1}{4}\tilde{\veps}^{(|x-y|\vee L_{k-2})^\chi} = \tfrac{1}{4}\veps^{q_{k-1}(|x-y|\vee L_{k-2})^\chi} < \tfrac{1}{4}\veps^{q_{k}(|x-y|\vee L_{k-2})^\chi}
\label{(1.38)}
\end{equation}
on the term $m=2$.

The terms $m>2$ may be handled inductively, since the bound just proven is sufficient to reproduce the argument when summing each pair $y_i,x_{i+1}$ in turn, and we obtain a bound $2^{-m}\tilde{\veps}^{(|x-y|\vee L_{k-2})^\chi}$ on the $m^{\mathrm{th}}$ term.

Now consider the term $m=1$, \textit{i.e.} the first term in (\ref{(1.33)}). This corresponds to a situation where $x,y$ are already contained in the same block on scale $k-1$. As with the $m\ge2$ terms, we have a bound given by $Q_{x,y}^{(k-1)}$ on that scale:
\begin{equation}
Q_{x,y}^{(k-1)} \le \veps^{q_{k-1}(|x-y|\vee L_{k-3})^\chi}.
\label{(1.39)}
\end{equation}
We need to improve this by replacing $L_{k-3}$ with $L_{k-2}$. For this, we can assume that $|x-y| \le L_{k-2}$. 

\tc{black}{First, suppose} the scale $k-1$ block containing $x,y$ is composed of 2 or more scale $k-2$ blocks. 
 \tc{black}{We obtain a bound which is more than sufficient,
because we receive two or more factors of $\veps^{q_{k-2} L_{k-4}^\chi}$ from (\ref{(1.32)}),
 which leads to $  \veps^{2 q_{k-2} L_{k-4}^\chi} < \veps^{q_{k-1}L_{k-2}^\chi}$.}
(Here we use the fact that $\chi = \tfrac{2}{9}$, so $2L_{k-4}^\chi = 2^{1-4/9}L_{k-2}^\chi$, and the excess over $L_{k-2}^\chi$ will allow us to control the sum over
relative positions of the two blocks as in the arguments for (\ref{(1.34)})-(\ref{(1.38)}) above.)
We should also consider the case $k=2$, for which we have at least a factor
$\veps^{1/5} \le \veps^{q_1L_0^\chi}$ for a single block on scale 1, as explained
in the proof of (\ref{(1.32)}) for $k=1$.
\tc{black}{}

On the other hand, suppose the $(k-1)$-block \tc{black}{containing $x,y$} consists of a single $(k-2)$-block.
If $k=3$, then $k-2 = 1$ and $\veps^{1/5} < \veps^{1/6} =
\veps^{q_2L_1^\chi}$, so the desired bound holds.
If not, then
at least one of two cases must hold: $n>L_{k-2}^{2\psi/5}$, or $\text{diam}(B) > L_{k-2}$. (Since the block does not join up with other blocks after two jumps in scale $L_{k-2}\rightarrow L_k$, its distance from other blocks must be at least
$4L_{k-2}$.  
As we are working on the case where $B$ is \textit{not} isolated on scale
$k-2$, we must violate one of the \textit{other} conditions of Definition \ref{def:isolated} -- provided $k-2 \ge 2$.)  For the case $n > L_{k-2}^{2\psi/5}$, we have by (\ref{(2.22f)}) a bound
\begin{equation}
\sum_{B_1\text{ containing }x\text{ and }y}  \tilde{P}^{(k-1)}(B_1)\veps^{-q_{k-1} n(B_1)} \exp\left(-|\log \veps|q_{k-1}L_{k-2}^{-\psi/15}n(B_1)\right)
\label{(2.31a)}
\end{equation}
on the first term of (\ref{(1.33)}). The sum over $B_1$ is bounded by (\ref{(1.39)}),
and we pick up an extra factor
\begin{equation}
\veps^{q_{k-1}L_{k-2}^{-\psi/15}n(B_1)} \le \veps^{q_{k-1}L_{k-2}^{\psi/3}} = \veps^{q_{k-1}L_{k-2}^{\chi}},
\label{(1.40)}
\end{equation}
which includes the desired improvement $L_{k-3}\rightarrow L_{k-2}$.
Finally, consider the case of a block on scale $k-2$ with diameter $>L_{k-2}$. If
this is the case, we must be able to find 
$x',y'$ in the block with $|x'-y'|>L_{k-2}$. Then we can use (\ref{(1.32)}), which becomes 
\begin{equation}
Q_{x',y'}^{(k-2)} \le \veps^{q_{k-2}|x'-y'|^\chi}\le \veps^{q_{k-2}L_{k-2}^\chi},
\label{(1.41)}
\end{equation}
and after summing over  $\le(2L_{k-2}+1)^d$ possibilities for $x'$ and for $y'$, given that $x,y$ are in the block, we obtain a bound $\veps^{q_{k-1}L_{k-2}^\chi}$, as in the other cases. (The drop $q_{k-2}\rightarrow q_{k-1}$ allows us to use the 
additional factor $\veps^{q_{k-2}L_{k-3}^{-\psi/20}L_{k-2}^\chi}$ to control the sums -- see (\ref{(1.42)}) below.)
Altogether, we have considered up to three cases for any particular $k$. Let us allow 
 $|x-y|>L_{k-2}$ again. We obtain the following bound on the $m=1$ term:
\begin{equation}
3\veps^{q_{k-1}(|x-y|\vee L_{k-2})^{\chi}} \le 3\veps^{q_k (|x-y|\vee L_{k-2})^{\chi}} 
\veps^{q_{k-1}L_{k-2}^{-\psi/20}L_{k-2}^{\chi}} < \tfrac{1}{4} \veps^{q_{k}(|x-y|\vee L_{k-2})^{\chi}}.
\label{(1.42)}
\end{equation}

We return to a consideration of $Q_{x,y}^{(k),2}$, which consists of terms from
(\ref{(1.31)}) that are isolated on scale $k-2$, which means that $k \ge 4$. In this
case we have from (\ref{(1.30)}) that $\tilde{P}^{(k)}(B) = \veps^{L_k^{3\psi/5}/3}$. 
We need to sum over all possible blocks of diameter $\le L_{k-2}$ and weight
by $\veps^{-q_kn}$ to obtain $Q_{x,y}^{(k),2}$. We obtain a bound
\begin{equation}
\veps^{L_{k-2}^{3\psi/5}/3}(2L_{k-1}+1)^{dn}\veps^{-q_kn} < \veps^{L_{k-2}^{3\psi/5}/4}
<\tfrac{1}{4}\veps^{q_kL_{k-2}^\chi} = \tfrac{1}{4}\veps^{q_k(|x-y|\vee L_{k-2})^\chi},
\label{(1.39a)}
\end{equation}
using $dn\log(2L_{k-2}+1) < c_dnL_{k-2}^{\psi/5} \le c_d L_{k-2}^{3\psi/5}$ and
$q_kn \le q_kL_{k-2}^{2\psi/5} < L_{k-2}^{3\psi/5}/12$ (and $q_k < \tfrac{1}{12}$ for $k \ge 4$, which can readily be checked from the recursion). Note that we are using
the Proposition \ref{prop:A} bound to control the sum over all substructure of $B$. 
This is because Proposition \ref{prop:A} cannot be used simultaneously for a block and for its subblocks, due to lack of independence. However, the bound for the whole block is much better than what had been obtained earlier for subblocks, and (\ref{(1.39a)}) demonstrates that it is adequate for our purposes here.

To complete the proof of the theorem, we gather the bounds (\ref{(1.42)}), (\ref{(1.39a)}) and the bound $2^{-m}\veps^{q_k(|x-y|\vee L_{k-2})^\chi}$ proven above on the $m^\text{th}$ term, $m \ge 2$. The desired bound (\ref{(1.32)}) follows, 
since $\tfrac{1}{4} + \tfrac{1}{4} + \sum_{m \ge 2}2^{-m} = 1$.\qed

We may now estimate $\calP_{x,y}^{(k)}$, the probability that $x$ and $y$ belong to the same resonant block on scale $k$.
\begin{corollary}
Given $\psi=\tfrac{2}{3}$, $\chi = \psi/3$, let $\veps$ be sufficiently small. Then there is a $q>0$ such that
\begin{equation}
\calP_{x,y}^{(k)} \le \veps^{q(|x-y|\vee L_{k-2})^{\chi}}.
\label{(1.43)}
\end{equation}
\label{cor:1'}
\end{corollary}
\textit{Proof.}
By construction, the probabilities $(3\sqrt{\veps})^{n(B)}$ and  
$\veps^{L_k^{3\psi/5}/3}$ in (\ref{(1.30)}) correspond to independent events, because whenever a block is resonant on a scale $k$, its probability bound from Proposition \ref{prop:A} replaces all previous probability factors for its subblocks. So any two subblocks with a factor $\veps^{L_k^{3\psi/5}/3}$ appearing in  (\ref{(1.30)}) are separated, ensuring independence. (Independence is a result of the truncation of the random walk expansion to form $\tilde{F}_E^{(k-1)}(B)$, including only graphs that remain within a distance $<L_{k-2}$ of $B$, so that there is no overlap between sets of variables $\mathbf{u}$ on which each resonance event depends.) At each length scale, isolated blocks must be resonant (otherwise they would be removed before forming blocks at the next scale) and hence can be associated with a probability factor  $\veps^{L_k^{3\psi/5}/3}$ from Proposition \ref{prop:A}.
Thus all the probability factors in (\ref{(1.30)}) correspond to independent events that must hold if $B$ is a component of $R^{(k)}$.
Hence $\tilde{P}^{(k)}(B)$ is a bound on the probability that $B$ is a component of $R^{(k)}$. 

Since the weights in (\ref{(1.31)})  are bounded below by 1, we see that (\ref{(1.32)}) becomes a bound on the sum  of $\tilde{P}^{(k)}(B)$ over $B$ containing $x,y$. The corollary follows, since $q_k > q > 0$.\qed

The underlying mechanisms at work in the proof of Theorem \ref{thm:1} are as follows. First, since we are only seeking fractional exponential decay, the procedure can tolerate giving up some fraction of the decay distance with each new length scale (which happens because of the gaps that were required between resonant components).
Smallness of the probability of a component comes in one of two ways. If a component is isolated, with small volume and diameter, the basic proposition on resonance probability can be used to bring in new smallness to control its greater positional freedom -- it gives an estimate that decays as a fractional exponential of the length scale. On the other hand, if the component's volume or diameter is large, or it is not too far from other components, the decay can be obtained from previous scales. We need to be careful not to give up too much volume decay for this purpose in each step, so that we can obtain bounds that are uniform in the step index $k$.

\subsection{Random Walk Expansion}\label{ssec:rwe}
The random walk expansion is based on the Schur complement at level $k$. We put
\begin{equation}
F_\lambda^{(k-1)} = \left( \begin{array}{cc}
A^{(k)} &B^{(k)} \\C^{(k)} & D^{(k)}  \end{array} \right),
\label{(1.44)}
\end{equation}
where the blocks are determined by the decomposition of $R^{(k-1)}$ into $R^{(k)}$ (upper-left block) and $R^{(k-1)}\setminus R^{(k)}$ (lower-right block). We may now define the $k^{\text{th}}$ Schur complement for $|\lambda - E| \le \veps_k /2$:
\begin{equation}
F_\lambda^{(k)} = 
A^{(k)} -B^{(k)} (D^{(k)}-\lambda)^{-1}  C^{(k)},
\label{(1.47)}
\end{equation}
Let us write
\begin{equation}
D^{(k)}=W^{(k)}+V^{(k)},
\label{(1.47a)}
\end{equation}
where $W^{(k)}$ is block diagonal, each block being $\tilde{F}_\lambda^{(k-1)}(B)$ for some $B$. This means that $V^{(k)}$ consists of the long graphs not included in $\tilde{F}_\lambda^{(k-1)}(B)$, producing matrix elements both within blocks and between blocks.

We show below in Theorem \ref{thm:3} that 
\begin{equation}
\|\tilde{F}_\lambda^{(k-1)}(B)-\tilde{F}_E^{(k-1)}(B)\| \le c_d\frac{\gamma}{\veps}|\lambda-E|,
\label{(1.47b)}
\end{equation}
which is less than $\veps_k/6$, because $|\lambda-E|\le \veps_k/2$. Since all the blocks of $R^{(k-1)}\setminus R^{(k)}$ are non-resonant,
\begin{equation}
\text{dist}\big(\text{spec}\,\tilde{F}_E^{(k-1)}(B),E\big) \ge \veps_k \equiv \veps^{L_k^\psi},
\label{(1.48)}
\end{equation}
and so
\begin{equation}
\|(W^{(k)}-\lambda)^{-1}\| \le 3\veps_k^{-1}.
\label{(1.49)}
\end{equation}
Hence, as in the first step, we may expand $(D^{(k)}-\lambda)^{-1}$ in a Neumann series:
\begin{equation}
[(D^{(k)}-\lambda)^{-1}]_{xy} = \sum_{g_k:x \rightarrow y}\, \prod_{i=1}^m \,[(W^{(k)}-\lambda)^{-1}]_{x_i\tilde{x}_i} \prod_{j=1}^{m-1} V^{(k)}_{\tilde{x_j},x_{j+1}.} 
\label{(1.50)}
\end{equation}
Here $g_k = \{x=x_1,\tilde{x}_1,x_2,\tilde{x}_2,\ldots,x_m,\tilde{x}_m=y\}$, with each $x_i,\tilde{x}_i$ in the same block. Note that $V^{(k)}_{x,y}$ is given by a sum of graphs contributing to $F_\lambda^{(k-1)} - \oplus_B \tilde{F}_\lambda^{(k-1)} (B)$, where
\begin{equation}
F_\lambda^{(k-1)} = 
A^{(k-1)} -B^{(k-1)} (D^{(k-1)}-\lambda)^{-1}  C^{(k-1)}.
\label{(1.51)}
\end{equation}
Thus, when it is useful, we may unwrap any matrix element into a sum of multigraphs, since each step at any level consists of a sum of graphs on the previous level. Thus, the matrices $A^{(k-1)},B^{(k-1)},C^{(k-1)},D^{(k-1)}$ in (\ref{(1.51)}) are themselves blocks of $F_\lambda^{(k-2)}$, which is given by $A^{(k-2)} -B^{(k-2)} (D^{(k-2)}-\lambda)^{-1}  C^{(k-2)}$. We may continue down to graphs at level 1.

We will also prove bounds on the matrices that generate the eigenfunctions. Recall from the fundamental lemma on the Schur complement that if $\varphi^{(k)}$ is an eigenvector of $F_\lambda^{(k)}$ with eigenvalue $\lambda$, then 
\begin{equation}
\varphi^{(k-1)}=\big(\varphi^{(k)}, -(D^{(k)}-\lambda)^{-1}  C^{(k)}\varphi^{(k)}\big)
\label{(1.52)}
\end{equation}
is an eigenvector of $F_\lambda^{(k-1)}$ with the same eigenvalue. This process may be repeated to extend the eigenvector $\varphi^{(k)}$ all the way out to the original lattice $\Lambda^*$, that is, to produce $\varphi^{(0)}$, an eigenvector of $H$. Let us write
\begin{equation}
\varphi^{(0)}=G_\lambda^{(k)}\varphi^{(k)},
\label{(1.53)}
\end{equation}
and then we may give a multigraph expansion for $G_\lambda^{(k)}$ in the same manner as was just described for $F_\lambda^{(k)}$. Indeed, the same operators $C^{(k)},D^{(k)}$ appear when unwrapping (\ref{(1.52)}). The matrix $G_\lambda^{(k)}$ can be thought of as half of $F_\lambda^{(k)}$ since it descends to the original lattice, but does not climb back up to $R^{(k)}$.

One may visualize the multigraph expansion on the original lattice as an ordinary random walk of nearest-neighbor steps, except that when it enters a block on level $k$, there is a matrix element of $(W^{(k)}-\lambda)^{-1}$ that produces an intra-block jump. This structure is present already at the first level, see (\ref{(1.17)}); but there 
\tc{black}{we had only single-position blocks.}
The larger blocks that appear in later steps create gaps in decay, since there is no decay from $x_i$ to $\tilde{x}_i$ in the following estimate for a walk passing through a block $B$ of $R^{(k-1)} \setminus R^{(k)}$:
\begin{equation}
\sum_{\tilde{x}_i} \left| [(W^{(k)}-\lambda)^{-1}]_{x_i\tilde{x}_i} \right| \le n(B)\cdot 3\veps_k^{-1} \le 3L_k^{2\psi/5}\veps^{-L_k^\psi}.
\label{(1.54)}
\end{equation}
Here we use (\ref{(1.49)}), and note that there are just $n(B)$ choices for $\tilde{x_i}$, given $x_i$. As the block is isolated on scale $k$, its volume is bounded by $L_k^{2\psi/5}$. The main issue with controlling the multigraph expansion is making sure that the gaps in decay and the large factors from the bound (\ref{(1.54)}) do not spoil too much the exponential decay proven in step 1. Our constructions ensure that there are gaps of size $4L_k$ between blocks at scale $k$. Hence, whatever rate of decay is proven at scale $k-1$, some small fraction will be lost at scale $k$. This is why we are led to fractional exponential decay estimates. Keep in mind that as far as the step $k$ random walk expansions are concerned, the configuration of blocks up to that level $(R^{(1)},R^{(2)},\ldots,R^{(k)})$ is fixed.

We now state our main theorem on graphical bounds. Let $\calS^{(k)}_{x,z,y}$ denote the sum of the absolute values of all multigraphs for $B^{(k)} (D^{(k)}-\lambda)^{-1}  C^{(k)}$ that go from $x$ to $y$ and that contain $z$. Here, $x,y$ are positions in $R^{(k)}$, and $z$ is a position in $R^{(k)}\setminus R^{(k-1)}$. We say that a multigraph contains $z$ if any of the blocks that it passes through contain $z$.
\begin{theorem}
With $\psi = \tfrac{2}{3}, \phi = \tfrac{1}{4}$, and  $\veps = \gamma^\phi$  sufficiently small, assume that $|\lambda-E| \le \veps_k/2$. Then
\begin{align}
\calS^{(k)}_{x,z,y} &\le \gamma^{  [(|x-z|+|z-y|)\vee L_k]^\psi }\cdot2^{-k} \label{(1.55)}
\\
\sum_{j \le k}\calS^{(j)}_{x,z,y} &\le \gamma^{  [(|x-z|+|z-y|)\vee 1]^\psi }.
\label{(1.55a)}
\end{align}
\label{thm:2}
\end{theorem}
\textit{Proof}. For the case $k=1$ we obtain
\begin{equation}
\calS^{(k)}_{x,z,y} \le \left(\frac{c_d\gamma}{\veps}\right)^{(|x-z|+|z-y|)\vee 2}\cdot \veps,
\label{(1.56)}
\end{equation}
as in the proof of (\ref{(1.20)}). Let $L = (|x-z|+|z-y|)\vee 1$, then one has a power $\tfrac{3}{4}(L\vee 2)+\tfrac{1}{4}$ of $\gamma$ in (\ref{(1.56)}). This is greater than the required power $L^{2/3}$ for (\ref{(1.55)}), as one can readily verify, checking the first few values of $L$ by hand. Hence for $\gamma$ sufficiently small, we obtain (\ref{(1.55)}).

\begin{figure}[h]
\centering
\includegraphics[width=.85\textwidth]{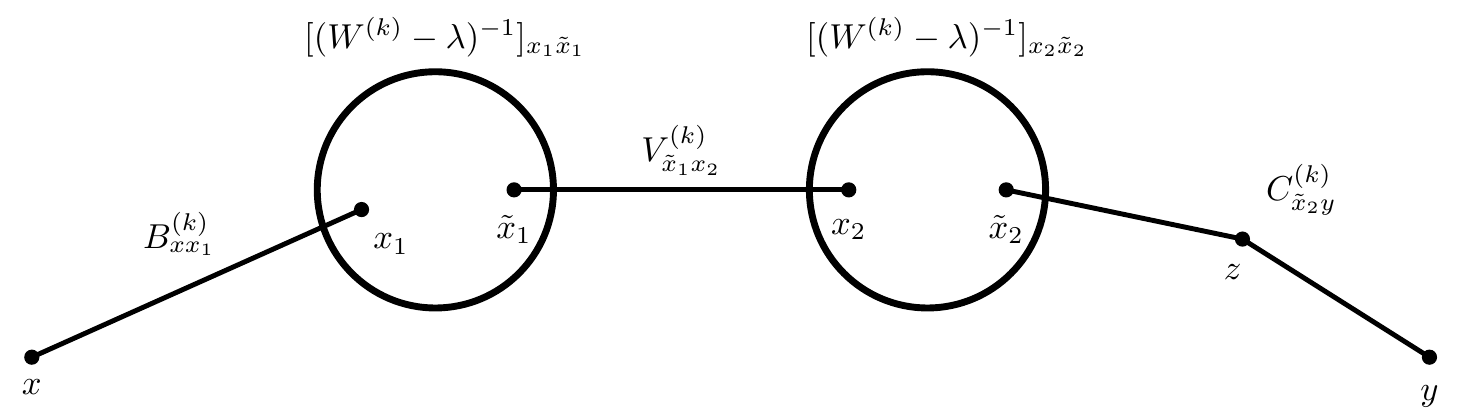}
\caption{A graph from $x$ to $y$ in the step $k$ random walk expansion} \label{fig:2}
\end{figure}

Let us now assume (\ref{(1.55)}) for $k-1$. Using (\ref{(1.49)}), one may expand $B^{(k)} (D^{(k)}-\lambda)^{-1}  C^{(k)}$ to obtain a graph connecting $x$ to $y$ traversing $m$ intermediate blocks. In the figure, each block has a matrix element of $(W^{(k)}-\lambda)^{-1}$ as in 
(\ref{(1.54)}). The extra point $z$ must lie in one of the sections of the graph (as in Figure \ref{fig:2}) or in one of the blocks. 
Because the blocks are isolated, and $x,y$ are in $R^{(k)}$, each section of the graph traverses a distance $\ge4L_k$, and the blocks themselves have diameter $\le L_k$. We sum over the pairs $(x_i,\tilde{x}_i)$ in turn in a manner similar to what was done for the probability estimate. In Figure \ref{fig:2}, for example, the sum over $x_i$ may be controlled with a combinatoric factor $c_d|x-x_i|^{d+1}$ and (\ref{(1.54)}) controls the sum over $\tilde{x}_1$ and results in a factor 
$3L_k^{2\psi/5}\veps^{-L_k^\psi}$. Applying the inductive hypothesis to $B^{(k)}$, $C^{(k)}$, and $V^{(k)}$ (which come from $F_\lambda^{(k-1)})$, we obtain a bound on the portion from $x$ to $x_2$:
\begin{equation}
\gamma^{|x-x_1|^\psi}c_d|x-x_1|^{d+1}\cdot 3L_k^{2\psi/5}\veps^{-L_k^\psi}\gamma^{|\tilde{x_1}-x_2|^\psi},
\label{(1.57)}
\end{equation}
noting that $|x-x_1| \equiv L_a \ge 4L_k$, and $|\tilde{x}_1-x_2| \equiv L_b \ge 4L_k$. We claim that
\begin{equation}
L_a^\psi - \phi L_k^\psi + L_b^\psi > (L_a+2L_k+L_b)^\psi \ge (|x-x_2|\vee 4L_k)^\psi.
\label{(1.58)}
\end{equation}
The first inequality may be reduced to the minimal case $L_a=L_b=4L_k$ by comparing derivatives of both sides. Then it becomes $4^\psi - \phi + 4^\psi > 10^\psi$, which is easy to verify for $\psi = \tfrac{2}{3}, \phi = \tfrac{1}{4}$. The second inequality is clear. Since  \tc{black}{the inequality is strict}
 in (\ref{(1.58)}), there is a sufficient leftover power of $\gamma$ to control the factors of $|x-x_1|$ and $L_k$ and obtain an overall bound
\begin{equation}
\gamma^{(|x-x_2|\vee 4L_k)^\psi}\cdot 4^{-k},
\label{(1.59)}
\end{equation}
which is sufficient to repeat the argument when summing over $(x_2,\tilde{x}_2)$ in the next block. In this manner we may proceed sequentially through a chain of $m$ blocks.

The point $z$ must lie either in one of the blocks or in one of the connecting graphs.
If a connecting graph is required to pass through $z$, then we may take, for example, $L_b = |\tilde{x}_1-z|+|z-x_2|$ and the argument goes through, achieving decay along a path that includes $z$. If $z$ is in a block, then the path from $x$ to $z$ to $y$ may entail a double traverse of the block, but this is covered by the factor of 2 in (\ref{(1.58)}). Furthermore, there is also the possibility of a self-line, where $V^{(k)}$ connects a block to itself. Such graphs contain a point $z'$ that is a distance $\ge L_{k-1}$ from $B$, so $|x-z'|+|z'-y|  \ge 2L_{k-1} = L_k$, and then summing
(\ref{(1.55)}) over $z'$ and $y$ gives a bound $c_d L_k^{d+1}L_k^d \gamma^{L_k^\psi}$ on the sum of all such graphs. Thus we may account for the self-lines with a factor 1.1, say, per block. Another factor per block controls the sum over which part of the graph contains $z$.
These considerations allow us to obtain from (\ref{(1.58)}) the desired decay distance $(|x-y|+|y-z|)\vee L_k$, along with a factor $3^{-km}$ for a chain of $m$ blocks from $x$ to $y$. Summing this over $m\ge1$ leads to the desired bound, (\ref{(1.55)}). The bound (\ref{(1.55a)}) on the sum of graphs on all scales follows immediately. This completes the proof. \qed

\begin{corollary} Under the same assumptions as Theorem \ref{thm:2},
\begin{equation}
\|F_\lambda^{(k)}  - \oplus_B \tilde{F}_\lambda^{(k)}(B)\| \le (c_d\gamma)^{L_k^\psi}.
\label{(1.59a)}
\end{equation}
\label{cor:2'}
\end{corollary}
\textit{Proof.} Graphs contributing to the difference go from $x$ to $y$ via a point $z$ such that $|x-z|\ge L_{k-1}$, $|y-z|\ge L_{k-1}$. We may bound the norm by estimating the maximum absolute row sum of the matrix. This means fixing $x$ and taking the sum over $z$ and $y$ of (\ref{(1.55a)}). A combinatoric factor $c_dL^{2d+2}$ controls the sums, where $L = |x-z|+|y-z| \ge 2L_{k-1} = L_k$, and (\ref{(1.59a)}) follows. \qed

We also need to control the difference $\tilde{F}_\lambda^{(k)}(B)-\tilde{F}_E^{(k)}(B)$ in norm, so that when isolated blocks are defined via the condition $\text{dist}\big(\text{spec}\,\tilde{F}_E^{(k-1)}(B),E\big) > \veps_k$, it is still safe to build the random walk expansion for the Schur complement with respect to $\lambda$.
\begin{theorem}
With $\psi = \tfrac{2}{3}, \phi = \tfrac{1}{4}$, and  $\veps = \gamma^\phi$  sufficiently small, assume that $|\lambda-E| \le \veps_k/2$. Then
\begin{equation}
\|\tilde{F}_\lambda^{(k)}(B)  - \tilde{F}_E^{(k)}(B)\| \le c_d\frac{\gamma}{\veps}|\lambda-E|.
\label{(1.60)}
\end{equation}
\label{thm:3}
\end{theorem}
\textit{Proof}. We have $F_\lambda^{(k)} = A^{(k)} - B^{(k)}(D^{(k)}-\lambda)^{-1}C^{(k)}$. In addition to the explicit appearance of $\lambda$, the matrices $A^{(k)},B^{(k)},C^{(k)},D^{(k)}$ depend on $\lambda$ for $k \ge 2$.  We already have control of the graphs contributing to these expressions by Theorem \ref{thm:2}. An inductive argument will allow us to control also the difference when we change $\lambda$ to $E$.

We begin by proving an analog of Theorem \ref{thm:2} to control the sum of differences of graphs, \textit{i.e.} each graph is evaluated at $\lambda$ and at $E$ and the difference taken. Let $\tilde\calS^{(k)}_{x,y}$ denote the sum of the absolute values of all difference multigraphs that contribute to $[B^{(k)}(D^{(k)}-\lambda)^{-1}C^{(k)}]_{xy}$.
We claim that
\begin{equation}
\tilde{\calS}^{(k)}_{x,y} \le \frac{1}{\veps_k}\gamma^{  (|x-y|\vee L_k)^\psi }\cdot2^{-k} |\lambda - E|,
\label{(1.61)}
\end{equation}
and hence that
\begin{equation}
\sum_{j \le k}\tilde{\calS}^{(j)}_{x,y} \le \frac{1}{\veps}\gamma^{  (|x-y|\vee 1)^\psi }|\lambda-E|.
\label{(1.61a)}
\end{equation}
Consider the case $k=1$. Redoing the proof of (\ref{(1.20)}) for differences, we obtain a sum of graphs wherein a difference
\begin{equation}
[(W^{(1)}-\lambda)^{-1} - (W^{(1)}-E)^{-1}]_{x_i\tilde{x}_i} = (\lambda-E)[(W^{(1)}-\lambda)^{-1} (W^{(1)}-E)^{-1}]_{x_i\tilde{x}_i}
\label{(1.62)}
\end{equation}
appears in place of the corresponding matrix element of $(W^{(1)}-\lambda)^{-1}$ or $(W^{(1)}-E)^{-1}$. Compared to what we had before, there is a combinatoric factor of $m$ for the choice of $i$, a factor $|\lambda - E|$, and an extra $1/\veps$ from the additional $(W^{(1)}-E)^{-1}$. We obtain in place of (\ref{(1.56)})
\begin{equation}
\tilde{\calS}^{(1)}_{x,y} \le \left(\frac{c_d\gamma}{\veps}\right)^{  (|x-y|\vee 2)} |\lambda - E| \le 
\frac{1}{2\veps}\gamma^{  (|x-y|\vee 1)^\psi }|\lambda-E|,
\label{(1.63)}
\end{equation}
which verifies (\ref{(1.61)}). For the second inequality of (\ref{(1.63)}), let $L = |x-y| \vee 1$. One can check that $\tfrac{3}{4}(L\vee 2)>L^{2/3} - \tfrac{1}{4}$, and then the claim may be verified by counting powers of $\gamma$.

For step $k$, we apply the difference operation to each term of (\ref{(1.50)}), and also to $B^{(k)},C^{(k)}$ in the expression $B^{(k)}(D^{(k)}-\lambda)^{-1}C^{(k)}$. Each matrix 
$W^{(k)},V^{(k)},B^{(k)},C^{(k)}$ is covered by (\ref{(1.61a)}), by induction, and this leads to an incremental factor of $\veps^{-1}|\lambda-E|$, compared to before. When we difference the explicit factors of $\lambda$ in (\ref{(1.50)}), we obtain as in (\ref{(1.62)}) a new factor of $(W^{(k)}-E)^{-1} |\lambda - E|$. This leads to an incremental factor $\veps_k^{-1}|\lambda-E|$, compared to before, coming from the bound $\|(W^{(k)}-E)^{-1}\| \le \veps_k^{-1}$. Thus in all cases, we get no worse than a factor $\veps_k^{-1}|\lambda-E|$. This completes the proof of (\ref{(1.61)}), (\ref{(1.61a)}).

Note that  (\ref{(1.61a)}) provides an estimate on the matrix elements of $F_\lambda^{(k)} - F_E^{(k)}$, so that
\begin{equation}
\sum_y \left|\big[F_\lambda^{(k)} - F_E^{(k)}\big]_{xy}\right| \le c_d \frac{\gamma}{\veps}|\lambda-E|,
\label{(1.64)}
\end{equation}
and hence
\begin{equation}
\|F_\lambda^{(k)} - F_E^{(k)}\| \le c_d \frac{\gamma}{\veps}|\lambda-E|,
\label{(1.65)}
\end{equation}
The same bound applies to $\|\tilde{F}_\lambda^{(k)}(B)- \tilde{F}_E^{(k)}(B)\|$, since in this case we are just looking at a subset of the collection of multigraphs (the ones that remain within a distance $<L_{k-1}$ of $B$).\qed
 
In order to prove decay of eigenfunctions, we need a bound similar to Theorem \ref{thm:2} on the sum of multigraph contributing to $G_\lambda^{(k)}$, the eigenfunction-generating kernel. Let $\calG_{x,y}^{(k)}$ denote the sum of the absolute values of all multigraphs for $G_\lambda^{(k)}$  -- see (\ref{(1.52)}),(\ref{(1.53)}) for its definition.

\begin{theorem}
With $\psi = \tfrac{2}{3}, \phi = \tfrac{1}{4}$, and  $\veps = \gamma^\phi$  sufficiently small, assume that $|\lambda-E| \le \veps_k/2$. Then
 \begin{align}
 \calG_{x,y}^{(k)} &\le \left(\frac{\gamma}{\veps}\right)^{(|x-y|\vee L_k)^\psi}\cdot2^{-k}
 \label{(1.66)}
 \\
\sum_{j \le k} \calG_{x,y}^{(j)} &\le \left(\frac{\gamma}{\veps}\right)^{(|x-y|\vee 1)^\psi}.
\label{(1.67)}
 \end{align}
 \label{thm:4}
 \end{theorem}
 Note that along with the graphical terms controlled by this theorem, $G_\lambda^{(k)}$ contains an identity matrix at the block for $R^{(k)}$, see (\ref{(1.52)}). The theorem only controls the nontrivial graphs in  $G_\lambda^{(k)}$.
 
\textit{Proof}. We have already established in Theorem \ref{thm:2} the needed bounds on  $D^{(k)},C^{(k)}$, and also $W^{(k)},V^{(k)}$, which appear in the random walk expansion (\ref{(1.50)}) for $(D^{(k)}-\lambda)^{-1}$. When estimating $(D^{(k)}-\lambda)^{-1}C^{(k)}$, we have a picture similar to Figure \ref{fig:2}, except without the $B^{(k)}$ line. In the proof of Theorem \ref{thm:2}, gaps in decay and factors $\veps_k^{-1}$ at blocks were controlled by (\ref{(1.58)}). There we had decay on either side of each block, which allowed us to prove uniform fractional exponential decay. Here we need to work with just the line on one side.
\begin{figure}[h]
\centering
\includegraphics[width=.45\textwidth]{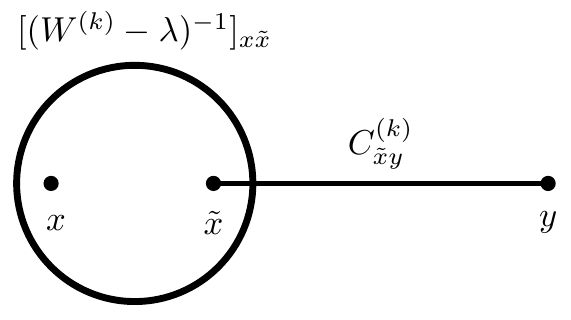}
\caption{A graph from the eigenfunction-generating kernel. } \label{fig:3}
\end{figure} 
Consider the situation in Figure \ref{fig:3}. Allowing for combinatoric factors as in (\ref{(1.57)}), we obtain a bound
\begin{equation}
3L_k^{\psi/2}\veps^{-L_k^\psi}\gamma^{|\tilde{x}-y|^\psi},
\label{(1.68)}
\end{equation}
with $|\tilde{x}-y| \equiv L_a \ge 4L_k$. We claim that
\begin{equation}
L_a^\psi  - \phi L_k^\psi > \tfrac{3}{4}(L_a + L_k)^\psi.
\label{(1.69)}
\end{equation}
The worst case for this inequality is when $L_a = 4L_k$, where it reduces to $4^\psi - \phi > \tfrac{3}{4}5^\psi$, which can easily be verified for $\phi = \tfrac{1}{4}, \psi = \tfrac{2}{3}$. Hence we may bound (\ref{(1.68)}) by
\begin{equation}
\left(\frac{\gamma}{\veps}\right)^{(|x-y|\vee L_k)^\psi}\cdot4^{-k}.
\label{(1.70)}
\end{equation}
Repeating this argument along a chain of blocks, and then along the chain of operators $\prod_j(-
(D^{(j)}-\lambda)^{-1}C^{(j)})$ that go into the definition of $G_\lambda^{(k)}$, we obtain (\ref{(1.66)}). The bound (\ref{(1.67)}) on the sum over scales follows immediately.\qed

\section{\tc{black}{Proofs of} Main Results}\label{sec:main}
We have established probabilistic estimates on resonant blocks, giving decay both in the spatial extent and in the volume of the block. Furthermore, we have obtained graphical estimates that imply decay of eigenfunctions away from resonant blocks. Let us put these tools to work by proving our main theorems. In the first section we prove our theorem on density of states. In Section \ref{ssec:efp}, we introduce an ``energy-following procedure'' that serves to construct every eigenstate through a process of successive approximation. In Section \ref{ssec:disc}, we prove the needed probabilistic bound on discriminants. In Section \ref{ssec:efc}, we complete the proof of fractional exponential decay of the eigenfunction correlator. In Section \ref{ssec:level}, we extend the method to complete the level-spacing result.
\subsection{Density of States}\label{ssec:dos}
\begin{theorem}
\label{thm:2.1}
There exists a constant $b>0$ such that for $\gamma$ sufficiently small the following is true. For any $0 < \delta \le \gamma^{1/4}$, any rectangle $\Lambda$, and any interval $I = [E-\delta/2,E+\delta/2]$, let $N(I)$ denote the number of eigenvalues of $H$ in $I$. Then
\begin{equation}
\mathbb{E}\,N(I) \le 4|\Lambda| \exp(-b|\log \gamma|^{2/3}|\log \delta|^{1/3}).
\label{(2.1)}
\end{equation}
\end{theorem}

The focus of this theorem is on the behavior as $\delta\rightarrow 0$, in particular we see that $\mathbb{E}N(I)$ vanishes faster than any power of $1/|\log \delta|$. If one considers instead a ``macroscopic'' interval with $\delta > \veps = \gamma^\phi = \gamma^{1/4}$, then the following bound holds:
\begin{equation}
\mathbb{E}\,N(I) \le 4\sqrt{\delta} |\Lambda|.
\label{(2.2)}
\end{equation}
This estimate may be obtained by a small modification of step 1. Define $R^{(1)}$ using the criterion that $t_x \in [E-\delta,E+\delta]$. Then as in (\ref{(1.11)}) we have a bound of $2\sqrt{\delta}$ on the probability that a position is in $R^{(1)}$. Weyl's inequality limits the movement of eigenvalues to $O(\gamma)$, hence $N(I)$ is bounded by $2|R^{(1)}|$. The expectation of $|R^{(1)}|$ is the sum on $x$ of the probability that $x \in R^{(1)}$, so (\ref{(2.2)}) follows from (\ref{(1.11)}).

\textit{Proof of Theorem \ref{thm:2.1}}. We follow the argument just given for (\ref{(2.2)}), but work with $R^{(k)}$ instead of $R^{(1)}$. Here $k$ is defined by the condition $\veps_{k+1} < \delta \le \veps_k$ (recall that $\veps_k \equiv \veps^{L_k^\psi}$). By assumption, $\delta \le \gamma^{1/4} = \veps = \veps_1$.
The probability that $x \in R^{(k)}$ is bounded by
\begin{equation}
\calP^{(k)}_{x,x} \le \veps^{qL_{k-2}^\chi} \le \veps^{\tilde{q}2^{k\chi}} = \veps^{\tilde{q}(2^{k\psi})^{1/3}},
\label{(2.3)}
\end{equation}
using (\ref{(1.43)}), $\tilde{q} \equiv q/8^\chi$, $\chi = \psi/3$. We may write this as 
\begin{align}
\exp(-\tilde{q}|\log \veps| |\log_\veps \veps_{k+1}|^{1/3}) &\le  \exp(-\tilde{q}|\log \veps| |\log_\veps \delta|^{1/3})  \nonumber
 \\ &= \exp(-\tilde{q}|\log \veps|^{2/3}|\log \delta|^{1/3}),
\label{(2.4)}
\end{align}
and after summing over $x$, we obtain (\ref{(2.1)}).\qed

\textit{Remark 1.} The bounds (\ref{(2.1)}),(\ref{(2.2)}) imply the following unified estimate, valid for $0 < \delta \le 1$:
\begin{equation}
\mathbb{E}\,N(I) \le 4|\Lambda| \exp(-b|\log (\gamma \vee \delta)|^{2/3}|\log \delta|^{1/3}).
\label{(2.1a)}
\end{equation}

\textit{Remark 2.} The fractional power of $|\log \delta|$ inside the exponential is a result of the discrepancy between $\chi$ (the probability exponent) and $\psi$ (the graphical exponent). The relation $\chi = \psi/3$ (or something similar) is necessitated ultimately by the $\tfrac{1}{2n}$ exponent in (\ref{(1.26)}), which leads to the 
volume limitation in Definition \ref{def:isolated}. This leads to limits on $\chi$ from  (\ref{(1.27)}) and  (\ref{(1.40)}), depending on how large $n$ is.

\subsection{Energy-Following Procedure}\label{ssec:efp}
 We will need a modified procedure that chooses energy windows to contain successive approximations to eigenvalues of $H$. Observe that the spectrum of $H$ is necessarily contained in the union of $2d\gamma$-neighborhoods of each $\pm t_x, x\in\Lambda$.
 (Note that $\|V\| = \|J\| \le 2d\gamma$, see (\ref{(1.3)}),(\ref{(1.4)}).) Let us choose one particular $x$ and put $E_1$ equal to one of the two eigenvalues $\pm t_x$ of $h_x$. Obviously $E_1$ depends on $x$ and on the choice of eigenvalue, but we suppress the dependence in the notation. We may then define the Schur complement $F^{(1)}_{E_1}$,
 as described in Section \ref{ssec:first}.
 
 Recall that blocks were formed by taking connected clusters of sites such that $\text{dist}(E_1,\{t_y, -t_y\}) \le \veps$ (based on nearest-neighbor connections). Note that if we wish to estimate the probability that $x,y$ are contained in the same block, as before there must be a sequence of distinct nearest-neighbor positions in $R^{(1)}$ connecting $x$ to $y$. The probability that any of these positions (other than $x$) is in $R^{(1)}$ is bounded by $3\sqrt{\veps}$, as before. By construction, $x$ is in $R^{(1)}$, so there is no small probability associated with the event $\{x \in R^{(1)}\}$. Thus we need to define $\tilde{\calP}_{x,y}^{(1)}$, the probability that $y$ belongs to the block based at $x$ (with a particular choice of $E_1$ taken as fixed). We have that
 \begin{equation}
\tilde{\calP}_{x,y}^{(1)} \le (12d\sqrt{\veps})^{|x-y|}.
\label{(2.5)}
\end{equation}
 Compared with (\ref{(1.12)}), we have one fewer factor of $12d\sqrt{\veps}$, for the reason just described. Graphical bounds such as (\ref{(1.20)}) are unaffected by the new procedure.
 
 To go on with the second step, we will need the truncated version of the Schur complement matrix. For $|\lambda - E_1| \le \veps/3$ we define as before $\tilde{F}_\lambda^{(1)}(B)$ by restricting the sum of graphs that defines $F_\lambda^{(1)}$, including only those that remain within a distance $<L_0 = \tfrac{1}{2}$ of $B$. Thus in this step, we consider no graphs of $B^{(1)}(D^{(1)}-\lambda)^{-1}C^{(1)}$. This means that $\tilde{F}_\lambda^{(1)}(B)$ does not actually depend on $\lambda$. But note that interaction terms in $V$ that hop between positions of $B$ are included in $\tilde{F}_\lambda^{(1)}(B)$. Since $\|V\| \le 2d \gamma$, the eigenvalues of $\tilde{F}_\lambda^{(1)}(B)$ are within $2d\gamma$ of the ``bare'' eigenvalues $\pm t_y, y \in B$. In a manner similar to what was done in the first step, we put $E_2$ equal to one of the eigenvalues of $\tilde{F}_{E_1}^{(1)}(B)$ in $[E_1 - \veps/3,E_1+\veps/3]$. There can be no more than $n(B)$ choices for $E_2$ given our initial choices of $(x, E_1)$.
 We will discuss the counting of choices $E_1, E_2,\ldots$ below; for now let us observe simply that in view of (\ref{(1.30)}), there is a rapid decay of probability with $n(B)$, so the choice of $E_2$ is under control. (There is a potential for redundancy here, because an eigenvalue may be reached via multiple initial choices of $(x,E_1)$; this is not a problem as long as every eigenvalue is covered at least once.)
 
 The process continues in the $k^{\text{th}}$ step in a similar fashion. At this point, there is a sequence of choices $x,E_1,\ldots,E_{k-1}$, and the associated increasing sequence of blocks containing $x$, which may be denoted $B_{x,1},\ldots,B_{x,k-1}$. We look at the spectrum of $\tilde{F}_\lambda^{(k-1)}(B_{x,k-1})$ in $[E_{k-1} - \veps_{k-1}/3, E_{k-1} + \veps_{k-1}/3]$, for $\lambda$ in this same interval. We look for solutions of the equation $\lambda \in \text{spec}\,\tilde{F}_\lambda^{(k-1)}(B_{x,k-1})$, as this closely approximates the exact equation $\lambda \in \text{spec}\, F_\lambda^{(k-1)}$ that determines eigenvalues of $H$ -- see the fundamental lemma (we discuss the existence of such solutions below). 
\tc{black}{Theorem \ref{thm:3} shows that the matrix $\tilde{F}_\lambda^{(k-1)}(B_{x,k-1})$ depends only weakly on $\lambda$: it obeys a Lipschitz condition with constant $c_d \gamma/\veps$. By Weyl's inequality, the same is true of its eigenvalues.} Hence one may sweep $\lambda$ through the interval above, and all the solutions to the equation $\lambda \in \text{spec}\,\tilde{F}_\lambda^{(k-1)}(B_{x,k-1})$ in 
 $[E_{k-1} - \veps_{k-1}/3, E_{k-1} + \veps_{k-1}/3]$ may be used as the next set of approximate eigenvalues. 
 We put $E_k$ equal to one such solution. (Note that by requiring $|E_k  - E_{k-1}| \le \veps_{k-1}/3$, we have that $|E_k-E_j| < \veps_j/2$ for $j<k$, because the sum of shifts $\veps_i/3$ for $j \le i < k$ is less than $\veps_{j}/2$. This ensures that we never leave the ``safe'' zone covered by Theorems \ref{thm:2}, \ref{thm:3}, and \ref{thm:4}.)
 Each choice may then serve as the central energy for the next Schur complement $F_{E_k}^{(k)}$ as defined in Section \ref{ssec:rwe}. It is important to note that when we shift to $E_k$, 
 we shift $F_{E_{k-1}}^{(j)}\rightarrow F_{E_k}^{(j)}$ for the random walk expansions at level $j<k$
 as well.
 We have the flexibility to do this because in step $j$, Theorem \ref{thm:2} controls graphical expansions constructed at any $\lambda$ such that $|\lambda - E_j| \le \veps_j/2$.

 There are some new aspects to the probability estimates for this ``energy-following'' procedure, as compared to the previous fixed-energy procedure. We need to adjust for the fact that by picking $E_k$ close to one of the solutions to $\lambda \in \text{spec}\,\tilde{F}_\lambda^{(k-1)}(B_{x,k-1})$, we lose the ability to obtain smallness of the probability of $B_{x,k-1}$, as $B_{x,k-1}$ is resonant to $E_k$ by construction.
 More precisely, there is no \textit{new} smallness in the probability estimate; when $B_{x,k-1}$ was formed in the previous step, any nontrivial structure was associated with smallness of probability (as we saw already in (\ref{(2.5)}) in step 1). Note, however, that once $B_{x,k-1}$ (the block containing $x$) is fixed, resonance probabilities for the \textit{other} blocks of $R^{(k-1)}$ involve random variables that are independent of those involved in defining $E_1,\ldots,E_k$ and $B_{x,k-1}$. This is why we always use the truncated matrices $\tilde{F}_\lambda^{(j)}(B_{x,j})$ to define the next set of central energies.
Due to this independence, the probability estimates for the blocks not containing $x$ (in particular Proposition \ref{prop:A}) are no different from before. Thus the Theorem \ref{thm:1} bound (\ref{(1.32)}) holds for blocks not containing $x$.

We prove a modified version of Theorem \ref{thm:1} that applies to the block containing $x$, for fixed choices of $E_1,\ldots,E_k$. Let $B_{x,k}$ denote the block of $x$ in step $k$, with $B_{x,0} \equiv \{x\}$. We introduce an altered version of the inductive definition (\ref{(1.30)}) that is suited to the energy-following procedure: 
\begin{align}
\hat{P}^{(1)}(B) &\equiv \begin{cases}(3\sqrt{\veps})^{n(B)-1}, & \text{if }x \in B ;\\
(3\sqrt{\veps})^{n(B)}, & \text{if }x \notin B
\end{cases} \nonumber
\\
\hat{P}^{(k)}(B) &\equiv 
\begin{cases}\veps^{L_{k-2}^{3\psi/5}/3}, & \text{if }B\text{ is isolated on scale }k-2\text{ and }x \notin B;
\\
\prod_{i=1}^{m'} \hat{P}^{(k-1)}(B_i), &\text{otherwise}.
\end{cases}
 \label{(2.6a)}
\end{align}
As before, $n(B)$ is the number of positions in $B$, and $B_1,\ldots,B_{m'}$ are the subcomponents of $B$ on scale $k-1$. 
It should be clear that this definition eliminates any factor corresponding to an event where $x$, or a block containing $x$, is resonant on any scale. As explained above, those events hold by construction in the energy-following procedure, so no smallness in probability is available. 
We also need the corresponding weighted sum of probability bounds, \textit{c.f.} (\ref{(1.31)}):
\begin{equation}
\hat{Q}^{(k)}_{x,y} \equiv \sum_{\substack{B\text{ containing }x\text{ and }y\\B_{x,k}\setminus B_{x,k-1}\ne\varnothing}} \hat{P}^{(k)}(B)\veps^{-q_k n(B)/2}.
\label{(2.6)}
\end{equation}
 \begin{theorem}
Given $\psi=\tfrac{2}{3}$, let $\chi = \psi/3$. Then for $\veps$ sufficiently small, and $q_1 \equiv \tfrac{1}{5}$, $q_k \equiv q_{k-1}(1-L_{k-2}^{-\psi/20})$, we have
\begin{equation}
\hat{Q}^{(k)}_{x,y} \le \veps^{q_k (|x-y|\vee L_{k-2})^\chi/2}.
\label{(2.7)}
\end{equation}
\label{thm:1'}
\end{theorem}

Note that no smallness of probability can be expected for a block that consists of the single site $x$; smallness arises only when \textit{other} blocks resonate with an energy from the block at $x$. But we see that for nontrivial blocks, the probability estimate is similar to Theorem \ref{thm:1}, but with a factor of $\tfrac{1}{2}$ in the exponents in (\ref{(2.6)}),(\ref{(2.7)}). Another difference is the requirement in (\ref{(2.6)}) that $B_{x,k} \setminus B_{x,k-1} \ne \varnothing$. This is needed because we may obtain the minimum decay length $L_{k-2}$ only if new resonant blocks have been added to $B_{x,k-1}$ in step $k$.

\textit{Proof of Theorem \ref{thm:1'}}. We follow the main steps of the proof of Theorem \ref{thm:1}. For the case $k=1$, note that $B_{x,1} \setminus B_{x,0} \ne \varnothing$ implies that $n>1$. Hence we may counterbalance the weighting by $\veps^{-q_1n/2}$ with the probability that $n-1$ positions other than $x$ are resonant, \textit{i.e.} $(3\sqrt{\veps})^{n-1} \le (3\sqrt{\veps})^{n/2}$ for $n>1$. Recalling that $q_1 = \tfrac{1}{5}$, we find that the sum of all trees with $n$ sites is bounded by $(2c_d\veps^{3/10})^{n/2}$. Then, since $n\ge|x-y|+1$, we obtain (\ref{(2.7)}) for $k=1$. The basis of this argument is that the missing factor of $3\sqrt{\veps}$ can be absorbed into the bound simply by halving the exponents, as was done in (\ref{(2.6)}),(\ref{(2.7)}).

When we move to the inductive step, we consider a tree graph on blocks, as in Figure \ref{fig:1}, and repeat the argument beginning with (\ref{(1.33)}). Only one of the scale $k-1$ blocks $B_1,\ldots,B_{m'}$ can contain $x$. We use the induction hypothesis (\ref{(2.7)}) if a block contains $x$, and the stronger bound (\ref{(1.32)}) from Theorem \ref{thm:1} otherwise.
However, there is no guarantee that $B_{x,k-1} \setminus B_{x,k-2} \ne \varnothing$. So we insert a partition of unity according to the value of $j$, the first scale at which $B_{x,j} = B_{x,k-1}$.
Then $B_{x,j} \setminus B_{x,j-1} \ne \varnothing$, and we have from (\ref{(2.7)}):
\begin{equation}
\sum_{j < k}\hat{Q}^{(j)}_{x,y} \le \sum_{j < k}\veps^{q_j (|x-y|\vee L_{j-2})^\chi/2} \le \veps^{q_{k}|x-y|^\chi/2},
\label{(2.8)}
\end{equation}
which is valid for $x \ne y$. (The sum over $j$ can be controlled via the decrease $q_j \rightarrow q_k$ in the exponent.) Compared with (\ref{(1.38)}), the exponent is halved and there is
no minimum decay length. Note that 
$B_{x,k} \setminus B_{x,k-1}\ne \varnothing$ implies that $m' >1$, so we may make a 
``halving argument'' as above to improve the deficient bound (\ref{(2.8)}) on the block containing $x$, at the expense of the others. Specifically, we use the inequality
\begin{equation}
\tfrac{1}{2}|x_1 - y_1|+|x_2 - y_2|\vee L_{k-3} \ge \tfrac{1}{2}(|x_1 - y_1|\vee L_{k-3}) + \tfrac{1}{2}(|x_2 - y_2|\vee L_{k-3}).
\label{(2.9)}
\end{equation}
Then it should be clear that we can apply the remaining arguments proving bounds such as (\ref{(1.38)}) on the terms $m\ge2$. We obtain an estimate 
$2^{-m}\veps^{q_k(|x-y|\vee L_{k-2})^\chi/2}$
on the $m^{\text{th}}$ term. 
Similar arguments will lead to the minimum decay length $L_{k-2}$ if $m=1$ and $m'>1$. The condition $B_{x,k} \setminus B_{x,k-1} \ne \varnothing$ eliminates the case $m=m'=1$. We obtain (\ref{(2.7)}) after summing over $m$.
\qed

As in Corollary \ref{cor:1'}, we may translate this theorem into an estimate on $\hat{\calP}_{x,y}^{(k)}$, the probability that $y$ belongs to the block of $x$ on scale $k$. Summing (\ref{(2.7)}) over $j$ as in (\ref{(2.8)}), we obtain the following result:
\begin{corollary}
Given $\psi=\tfrac{2}{3}$, $\chi = \psi/3$, let $\veps$ be sufficiently small. Then there is a $q>0$ such that
\begin{equation}
\hat{\calP}_{x,y}^{(k)} \le \veps^{q|x-y|^{\chi}/2}.
\label{(2.9.1)}
\end{equation}
\label{cor:2''}
\end{corollary}

\subsection{Discriminants}\label{ssec:disc}
 The following result is important for controlling the energy-following procedure, and it will also enter into the proof of our main level-spacing theorem for $H$. It is an analog of (\ref{(1.27)}) that applies to discriminants, instead of determinants. 
 It will be used to control the probability that the level spacing of a block is less than $\delta$.
 Let us put
\begin{equation}
\Gamma(\mathbf{u}) = \text{disc}\left(\tilde{F}^{(k)}_\lambda(B)\right),
\label{(2.9a)}
\end{equation}
where $\text{disc}(M) = \prod_{i<j}(\lambda_i-\lambda_j)^2$ is the discriminant of a matrix $M$ with eigenvalues $\lambda_1,\ldots,\lambda_N$. It is well-known that $\text{disc}(M)$ is a homogeneous polynomial of degree $N(N-1)$ in the entries $M_{ij}$.
\begin{proposition}
Let $B$ be a block of volume $n$. Consider the discriminant of the Schur complement matrix $\tilde{F}_\lambda^{(k)}(B)$. For any $\delta>0$,
\begin{equation}
P\left(|\Gamma(\mathbf{u})| \le \delta^23^{4n^2}\right) \le \delta^{1/(2n^2)}\cdot 4(4n)^{n+1}.
\label{(2.9b)}
\end{equation}
Furthermore,
\begin{equation}
P\left(\min_{i<j}|\lambda_i-\lambda_j| < \delta\right) \le \delta^{1/(2n^2)}\cdot 4(4n)^{n+1},
\label{(2.9c)}
\end{equation}
where $\lambda_1,\ldots,\lambda_{2n}$ are the eigenvalues of $\tilde{F}_\lambda^{(k)}(B)$.
\label{prop:B}
\end{proposition}
\textit{Proof}. As explained in the proof of (\ref{(1.27)}), the entries of $\tilde{F}_\lambda^{(k)}(B)$ are equal to those of $H$, plus terms of size $O(\gamma^2/\veps)$ that are independent of $\mathbf{u} \equiv \{u_x\}_{x\in B}$. Hence $\Gamma(\mathbf{u})$ is a polynomial of degree $\le4n^2$ in $\mathbf{u}$. (Here $n$ is the number of positions in $B$, so that $N=2n$.)

We apply the Brudnyi-Ganzburg inequality (\ref{(1.24)}) to $\Gamma(\mathbf{u})$. We need to find a point in parameter space where a lower bound on $\Gamma(\mathbf{u})$ can be proven. Take $\mathbf{u}_1 = (1,2,\ldots,n)$. Then the eigenvalues are 
close to $(\pm\sqrt{2},\pm\sqrt{5},\ldots,\pm\sqrt{1+n^2})$. The minimum gap is $\sqrt{5}-\sqrt{2} \approx .82$, and allowing for eigenvalue movement $O(\gamma^2/\veps)$, we obtain a crude lower bound
\begin{equation}
\Gamma(\mathbf{u}_1) \ge (4/5)^{4n^2}.
\label{(2.9d)}
\end{equation}
Then, following the arguments for (\ref{(1.25)})-(\ref{(1.27)}), we put $U = [-n,n]^n$, $\kappa = 4n^2$, and
\begin{equation}
\omega = \{\mathbf{u}:\, \Gamma(\mathbf{u}) \le \delta^23^{4n^2}\}.
\label{(2.9e)}
\end{equation}
Since $\sup_U |\Gamma(\mathbf{u})| \ge (4/5)^{4n^2}$ and $|U| = (2n)^n$, we obtain
\begin{equation}
|\omega| \le \left[(5/4)^{4n^2}\delta^23^{4n^2}\right]^{1/(4n^2)}\cdot 4n(2n)^n \le \tfrac{15}{4}\delta^{1/(2n^2)}\cdot 2(2n)^{n+1}.
\label{(2.9f)}
\end{equation}
A factor $2^n$ converts this into a probability, and we obtain (\ref{(2.9b)}). Now, restricting
to the support [-1,1] of the random variables $u_i$, the spectrum is contained in
$[-\tfrac{3}{2},\tfrac{3}{2}]$, so all eigenvalue differences are bounded by 3. Hence, if 
$\min_{i<j} |\lambda_i-\lambda_j| < \delta$, then $\Gamma(\mathbf{u}) < \delta^23^{4n^2}$.
Thus (\ref{(2.9b)}) implies (\ref{(2.9c)}) and the proof is complete.\qed

Let us relate this result to our energy-following procedure, in the context of a particular choice of $E_1,\ldots,E_k$.
\begin{definition}
A block $B$ of volume $n$ is said to be \textbf{autoresonant on scale k} if $n\le L_k^{\psi/4}$, $\text{diam}(B) \le L_k$, and
\begin{equation}
\min_{i<j} |\lambda_i-\lambda_j| < \veps_k \equiv \veps^{L_k^\psi},
\label{(2.9g)}
\end{equation}
where $\lambda_i, i = 1,\ldots,2n$ are the eigenvalues of $\tilde{F}^{(k)}_{E_k}(B)$.
\label{def:A}
\end{definition}
\begin{corollary}
Given $x$, $E_1,\ldots,E_k$ in the energy-following procedure, the probability that \textit{any} block containing $x$ is autoresonant on scale $k$ is bounded by $\veps^{L_k^{\psi/3}/2}$, for $\gamma$ sufficiently small.
\label{cor:B'}
\end{corollary}
\textit{Proof}. Given the restrictions on diameter and volume, the number of possible blocks $B$ is bounded by
\begin{equation}
(2L_k)^{dn} \le e^{dL_k^{\psi/4}\log 2L_k}.
\label{(2.9h)}
\end{equation}
For a given $B$, (\ref{(2.9c)}) shows that the probability is bounded by
\begin{equation}
\veps_k^{1/(2n^2)}\cdot4(4n)^{n+1} \le \veps^{L_k^{\psi/2}/2}\cdot4(4n)^{n+1}.
\label{(2.9i)}
\end{equation}
A combinatoric factor $2^n$ allows us to fix $n$. The corollary follows by taking the product of these bounds.\qed

 \subsection{Eigenfunction Correlators}\label{ssec:efc}
 We wish to prove fractional exponential decay of $\mathbb{E} \sum_\alpha |\varphi_\alpha(y)\varphi_\alpha(z)|$, which is  a strong form of localization. One should be able to prove that $\sum_\alpha |\varphi_\alpha(y)\varphi_\alpha(z)|$ is exponentially small, except on a set whose probability tends to zero rapidly with $|y-z|$. One would introduce connectedness conditions on a rapidly growing sequence of length scales as in \cite{Frohlich1983}, for example. Here, we focus on decay of the averaged eigenfunction correlator, which in our procedure can decay no more rapidly than do the probability estimates, \textit{i.e.} as a fractional exponential.
 
 Note that the eigenvalues of $H$ are simple, with probability 1. This can be seen by noting that the discriminant is nonzero, for a particular set of well-spaced $u$'s (of order $|\Lambda|$), because the eigenvalues remain separated after turning on parts of $H$ that connect different positions. The discriminant is a polynomial in $\mathbf{u}$, so it cannot vanish on a set of positive measure without being identically zero.
 
 \begin{theorem}
\label{thm:2.2}
Let $\chi = \tfrac{2}{9}$. There exists a constant $\bar{q}>0$ such that for $\gamma$ sufficiently small, the eigenfunction correlator satisfies 
\begin{equation}
\mathbb{E} \sum_\alpha |\varphi_\alpha(y)\varphi_\alpha(z)| \le 4\gamma^{\bar{q}|y-z|^\chi},
\label{(2.4a)}
\end{equation}
for any rectangle $\Lambda$.
\end{theorem}
\textit{Proof}. There are three parts of the proof. In the first part, we relate the sum on $\alpha$ in (\ref{(2.4a)}) to the energy-following procedure. 
We will establish that every eigenvalue $\lambda_0$ of $H$ can be reached via the energy-following procedure for at least one set of choices for $x,E_1,E_2,\ldots$. Specifically, there is a set of choices such that $E_k = \lambda_0$ for all $k$ sufficiently large (depending on $\lambda_0$).
Initially, we will allow ourselves the freedom to pick from amongst all the sites or blocks that have spectrum close to $\lambda_0$ at each scale. Then, we will show that there is at least one \textit{good} choice of $x$ such that the procedure converges to $\lambda_0$ when working with the block containing $x$ at each scale.

We proceed step by step. In the first step, consider the set of positions $x$ such that $\pm t_x$ is within $\veps/3$ of $\lambda_0$. Clearly, this set must be nonempty, because otherwise there would be no spectrum of $H_0$ within $\veps/3$ of $\lambda_0$, and $H = H_0+V$ with $\|V\| \le 2d\gamma$. For any of these choices of $x$, we can choose $E_1 = \pm t_x$ with $|E_1  - \lambda_0| \le \veps/3$. We then construct $R^{(1)}$ and $F_\lambda^{(1)}$ for $|\lambda-E_1|\le \veps/2$. By the fundamental lemma, $\lambda_0 \in \text{spec}\,F_{\lambda_0}^{(1)}$.

Now consider what happens after having made choices for $x_1,E_1,E_2,\ldots,E_k$ such that $E_j \in \text{spec}\,\tilde{F}^{(j+1)}_{E_j}(B)$ and $|E_j - \lambda_0| \le \veps_j/3$ for $j \le k$. Note that 
 $|E_j - \lambda_0| \le \veps_j/3$ implies that $\lambda_0 \in \text{spec}\,F^{(j)}_{\lambda_0}$, by repeated applications of the fundamental lemma. Let us truncate
 $F^{(k)}_{\lambda_0} \rightarrow \oplus_B\tilde{F}^{(k)}_{\lambda_0}(B)$. The difference has 
 norm $\le (c_d\gamma)^{L_k^\psi} \le \veps_k^3$, by Corollary \ref{cor:2'}. Therefore, at least 
 one block has spectrum close to $\lambda_0$, in the sense that $\text{dist}
 \big(\lambda_0,\text{spec}\,\tilde{F}^{(k)}_{\lambda_0}(B)\big) \le \veps_k^3$. For any such block, we wish to choose $E_{k+1}$ to be a solution in
 $[\lambda_0 - \veps_{k+1}/3,\lambda_0 + \veps_{k+1}/3]$ to the equation $\lambda \in
 \text{spec}\,\tilde{F}^{(k)}_{\lambda}(B)$. 
 Let us order the eigenvalues of $\tilde{F}^{(k)}_{\lambda}(B)$ as $\lambda_1(\lambda) \le \cdots 
 \le \lambda_{n(B)}(\lambda)$. For some $p$, $|\lambda_p(\lambda_0)-\lambda_0| \le \veps_k^3$. By the Lipschitz continuity of $\tilde{F}^{(k)}_{\lambda}(B)$ in $\lambda$ (Theorem \ref{thm:3}) and Weyl's inequality, the interval  $[\lambda_0 - 2\veps_{k+1}/3,\lambda_0 + 2\veps_{k+1}/3]$ maps contractively into itself under the map $\lambda \rightarrow \lambda_p(\lambda)$. By the
 contraction mapping principle, 
there is a solution $E_{k+1}$ to the equation $E_{k+1}=\lambda_p(E_{k+1})$ that satisfies 
 $|E_{k+1}-\lambda_0| \le 2\veps_k^3 \le \veps_{k+1}/3$.
 This may be used to construct $R^{(k+1)}$ and $F_{E_{k+1}}^{(k+1)}$, and the procedure continues.
 
 For $L_k > \text{diam }(\Lambda)$, there can be at most one block, and $\tilde{F}^{(k)}_{\lambda}(B) = F^{(k)}_{\lambda}$. Also, as explained above, all eigenvalues are simple, so eventually there will be exactly one solution to $\lambda \in \text{spec}\,F_\lambda^{(k)}$
 in  $[\lambda_0 - \veps_{k+1}/3,\lambda_0 + \veps_{k+1}/3]$, namely $\lambda_0$ itself. Thus we have established convergence of the procedure -- at least if one is allowed the freedom to pick any of the sites/blocks that have spectrum close to $\lambda_0$ at each scale. However, we would like a somewhat stronger statement:  that we can reach $\lambda_0$ starting at some $x$ and continuing through a sequence of blocks $B_{x,j}$ containing $x$. To see this, start at scale $k$ such that $\tilde{F}^{(k)}_{\lambda}(B) = F^{(k)}_{\lambda}$
 and $\lambda_0 \in \text{spec}\,F_{\lambda_0}^{(k)}$. As we proceed downward in scale, we can always conclude that at least one subblock is resonant with $\lambda_0$ (in the sense that $E_{j+1} \in \text{spec}\,\tilde{F}_{E_{j+1}}^{(j)}(B)$ and $|E_{j+1} - \lambda_0| < 2\veps_{j+1}^3$). The argument is the same as the one given above, except that $\tilde{F}_{E_{j+1}}^{(j)}$ (rather than $F_{\lambda_0}^{(j)}$) is the matrix that is truncated to its subblocks.
  Then at least one subblock has $E_{j} \in \text{spec}\,\tilde{F}_{E_{j}}^{(j-1)}$ with $|E_j  - E_{j+1}| \le \veps_j^3$. Proceeding down to the level of individual positions, we find one or more choices of $x$ that can be used as a starting point for the energy-following procedure, as claimed above. This completes the first part of the proof.
  
Let us define $N_{x,y}$ to be the number of eigenvalues of $H$ that can be reached via the energy-following procedure starting at $x$, with a resonant region that includes $y$.
In the second part of the proof, we obtain bounds on the expectation of $N_{x,y}$ -- see  (\ref{(2.13)}),(\ref{(2.14)}) below. For each $k$, the procedure involves a choice of eigenvalue $E_k$, which we will sum over so as to account for every eigenvalue that can be reached starting at $x$. Once $E_k$ is selected, there is a sum over $B_{x,k}$, the block containing $x$ (along with an associated probability -- the block sum is a partition of unity in the probability space). 
  We may define a ``stopping scale'' $\underline{k}$ as the first $k$ such that all of the following conditions hold:
  \begin{enumerate}
  \item{$B_{x,k} = B_{x,\bar{k}}$, where $\bar{k}$ is the smallest integer with $L_{\bar{k}} > \text{diam}(\Lambda)$. In other words, $B_{x,k}$ has reached its maximum extent.}
  \item{$n(B_{x,k}) \le L_k^{\psi/4}$.}
  \item{$\text{diam}(B_{x,k}) \le L_k$.}
  \item{$B_{x,k}$ is not autoresonant on scale $k$.}
  \end{enumerate}
 After scale $\underline{k}$ is reached, there is no further enlargement of $B_{x,k}$. Furthermore, conditions 2-4 imply that the eigenvalues of $\tilde{F}_{E_{k}}^{(k)}(B_{x,k})$ obey (\ref{(2.9g)}), 
 \textit{i.e.} they have a minimum spacing $\veps_k$. This means that there is only one choice for $E_{\underline{k}+1},E_{\underline{k} +2},\ldots$ (because for $k \ge \underline{k}$ the intervals $[E_k - \veps_{k+1}/3,E_k + \veps_{k+1}/3]$ can contain no more than one solution to the equation $E_{k+1} \in \text{spec}\,\tilde{F}_{E_{k+1}}^{(k)}(B_{x,k})$.)
 As a result, we may bound the entirety of the choices of $E_1,E_2,\ldots$ by 
 $[2n(B_{x,\underline{k}})]^{\underline{k}}$. (In step $k$, there are no more than $2n(B_{x,k})$ eigenvalues of $\tilde{F}_{E_{k}}^{(k)}(B_{x,k})$, hence no more than $2n(B_{x,k}) \le 2n(B_{x,\underline{k}})$
 solutions to $E_{k+1} \in \text{spec}\,\tilde{F}_{E_{k+1}}^{(k)}(B_{x,k})$.)
 For a given set of choices $E_1,E_2,\ldots$, the probability that $B_{x,\underline{k}}$ has
 size $n$ and contains $y$ is controlled by Theorem \ref{thm:1'}. Likewise, Proposition 
 \ref{prop:B} controls the probability of autoresonance.
 
 As in the proof of Corollary \ref{cor:1'}, we know that $\hat{P}^{(k)}(B)$ is a bound on the probability that $B_{x,k} = B$, \textit{i.e.} that $B$ is the block containing $x$ at step $k$ (for a given sequence of energies $E_1,E_2,...$ for the energy-following procedure). Note that if $\underline{k}$ is the first scale where conditions 1-4 all hold, then at least one condition fails when $k = \tilde{k} \equiv \underline{k}-1$.
 Thus we consider four cases, and bound
 \begin{equation}
\mathbb{E}\,N_{x,y} \le \mathbb{E}\,N^{(1)}_{x,y}+\mathbb{E}\,N^{(2)}_{x,y}+\mathbb{E}\,N^{(3)}_{x,y}+\mathbb{E}\,N^{(4)}_{x,y},
\label{(2.10)}
\end{equation}
 where each term represents the case where the corresponding condition breaks down for $k = \tilde{k}$. For case 1, this means that $B_{x,\underline{k}}\setminus B_{x,\underline{k}-1} \ne \varnothing$. Let us assume $x \ne y$ for the moment.
 Then we may estimate
 \begin{align}
\mathbb{E}\,N^{(1)}_{x,y} &\le \sum_{\underline{k}=1}^{\infty}\, \sum_{\substack{B\text{ containing }x\text{ and }y\\B_{x,\underline{k}}\setminus B_{x,\underline{k}-1}\ne\varnothing}}\,[2n(B)]^{\underline{k}-1} \hat{P}^{(\underline{k})}(B) 
\nonumber \\
&\le \sum_{\underline{k}=1}^{\infty}\, \sum_{\substack{B\text{ containing }x\text{ and }y\\B_{x,\underline{k}}\setminus B_{x,\underline{k}-1}\ne\varnothing}}\,(\underline{k}-1)!\,e^{2n(B)} \hat{P}^{(\underline{k})}(B)
\nonumber \\
&\le \sum_{\underline{k}=1}^\infty \underline{k}!\, \hat{Q}_{x,y}^{(\underline{k})} \le  \sum_{\underline{k}=1}^\infty \underline{k}!\, 
\veps^{q(|x-y|\vee L_{\underline{k}-2})^\chi/2},
\label{(2.11)}
\end{align}
 since for $\gamma$ small $e^{2n} \le \veps^{-qn/2}$, \textit{c.f.} (\ref{(2.6)}).
 The last inequality is from Theorem \ref{thm:1'} (recall that $q_k \searrow q$ as $k \rightarrow \infty$).
 Noting that the exponential growth of $L_{k-2}^\chi$ with $k$ dominates $k\log k$ from Stirling's formula, we have that
 \begin{equation}
\mathbb{E}\,N^{(1)}_{x,y}  \le \veps^{q|x-y|^\chi/3}.
\label{(2.12)}
\end{equation}
 For case 2, with $\tilde{k} = \underline{k}-1$ we bound $(2n(B))^{\tilde{k}}$ by $\tilde{k}!e^{2n(B)}$ as above. Then since $n(B_{x,\tilde{k}}) > L_{\tilde{k}}^{\psi/4}$, we may write
 \begin{align}
\mathbb{E}\,N^{(2)}_{x,y} &\le \sum_{\tilde{k}=1}^{\infty}\, \sum_{\substack{B\text{ containing }x\text{ and }y\\n(B) > L_{\tilde{k}}^{\psi/4}}}\,[2n(B)]^{\tilde{k}-1} \hat{P}^{(\tilde{k})}(B) 
\nonumber \\
&\le \sum_{\tilde{k}=1}^\infty \tilde{k}!\, \veps^{qL_{\tilde{k}}^{\psi/4}/3} \sum_{j\le \tilde{k}}\hat{Q}_{x,y}^{(j)}  \le  
\veps^{q|x-y|^\chi/3},
\label{(2.12a)}
\end{align}
 where we have exploited the decay in $n$ that is built into (\ref{(2.6)}) and applied (\ref{(2.8)}). For case 3, we have that $\text{diam}(B_{x,\tilde{k}}) > L_{\tilde{k}}$, so
$B_{x,\tilde{k}}$ must contain a point $z$ with $|z-x| > L_{\tilde{k}}$ (take $z=y$ if $|x-y|>L_{\tilde{k}}$). Then (after absorbing the factor $e^{2n}$ into $\hat{Q}$ as above)
\begin{align}
\mathbb{E}\,N^{(3)}_{x,y} &\le \sum_{\tilde{k}:\,L_{\tilde{k}} > |x-y|}\tilde{k}! \sum_{j\le \tilde{k}}\, \sum_{z:\,|z-x|>L_{\tilde{k}}}\hat{Q}_{x,z}^{(j)} 
+  \sum_{\tilde{k}:\,L_{\tilde{k}} \le |x-y|}\tilde{k}! \sum_{j\le \tilde{k}} \hat{Q}_{x,y}^{(j)} 
\nonumber \\
&\le \sum_{\tilde{k}=1}^\infty \tilde{k}!\, \veps^{q(|x-y|\vee L_{\tilde{k}})^{\chi}/2.5} \le  
\veps^{q|x-y|^\chi/3}.
\label{(2.12b)}
\end{align}
Here we have used (\ref{(2.8)}) again, and since in either term there is a minimum decay length $L_{\tilde{k}}$, we may use some of that decay to control the sums over $z$ and $\tilde{k}$. Finally, in case 4, $B_{x,\tilde{k}}$ is autoresonant on scale $\tilde{k}$, which implies that $|x-y| \le L_{\tilde{k}}$ and $n \le L_{\tilde{k}}^{\psi/4}$ (here $n = n(B_{x,\tilde{k}}))$. Therefore, $(2n)^{\tilde{k}} \le \exp\big(\tilde{k}\log (2n)\big) \le \exp(\tilde{k}^2)$. We apply Corollary \ref{cor:B'} to bound the probability, and then
\begin{equation}
\mathbb{E}\,N^{(4)}_{x,y} \le \sum_{\tilde{k}:\,L_{\tilde{k}} \ge |x-y|}e^{\tilde{k}^2} \veps^{L_{\tilde{k}}^{\psi/3}/2} \le \veps^{{|x-y|}^{\psi/3}/3} = \veps^{{|x-y|}^{\chi}/3}.
\label{(2.12c)}
\end{equation}

Combining the bounds on the four cases, we obtain
\begin{equation}
\mathbb{E}\,N_{x,y} \le  \veps^{{q|x-y|}^{\chi}/4}, \text{ if } x \ne y.
\label{(2.13)}
\end{equation}
If $x=y$, the above bounds apply when $\tilde{k} \ge 1$. But we also need to consider the case $\tilde{k}=0$, which means that the block containing $x$ contains no other points. Then there are only two eigenvalues that can be reached, and so
\begin{equation}
\mathbb{E}\,N_{x,x} \le 3.
\label{(2.14)}
\end{equation}

The third part completes the proof of Theorem \ref{thm:2.2}.
We follow the proof of (\ref{(2.13)}),(\ref{(2.14)}), except instead of simply counting eigenvalues, we weight them by $|\varphi_\alpha(y)\varphi_\alpha(z)|$. From (\ref{(1.53)}) we know that any eigenfunction $\varphi$ of $H$ can be written as $G_\lambda^{(k)}\varphi^{(k)}$, where $\varphi^{(k)}$ is an eigenvector of $F_\lambda^{(k)}$ with eigenvalue $\lambda$. Here we may take $k=\underline{k}\vee\bar{k}$, which ensures that $E_k$ is resolved and that $\tilde{F}_{E_k}^{(k)}(B) = F_{E_k}^{(k)}$, so that $E_k$ is an eigenvalue of $H$.

Let $N_{x,y,z}$ denote the number of eigenvalues of $H$ that can be reached via the energy-following procedure starting at $x$, and whose resonant region includes $y,z$.
Then clearly
\begin{equation}
\mathbb{E}\,N_{x,y,z} \le \mathbb{E}\,N_{x,y} \text{ and } \mathbb{E}\,N_{x,y,z} \le \mathbb{E}\,N_{x,z}.
\label{(2.16)}
\end{equation}
Taking the geometric mean, we obtain
\begin{equation}
\mathbb{E}\,N_{x,y,z} \le \big(\mathbb{E}\,N_{x,y} \big)^{1/2}\big(  \mathbb{E}\,N_{x,z}\big)^{1/2} \le 3\veps^{q|x-y|^\chi/8}\veps^{q|x-z|^\chi/8}.
\label{(2.17)}
\end{equation}
where we have applied the bounds (\ref{(2.13)}),(\ref{(2.14)}).

Theorem \ref{thm:4} implies that
\begin{align}
\sum_{\alpha} |\varphi_\alpha(y)\varphi_\alpha(z)| &= \sum_\alpha 
\left|\big(G_{\lambda_\alpha}^{(k)}\varphi_\alpha^{(k)}\big)(y)\big(G_{\lambda_\alpha}^{(k)}\varphi_\alpha^{(k)}\big)(z)\right|
\nonumber\\
&\le \sum_{x,y_1,z_1} 
\left(\delta_{yy_1} + \left(\frac{\gamma}{\veps}\right)^{|y-y_1|^\psi}\right)
\left(\delta_{zz_1} + \left(\frac{\gamma}{\veps}\right)^{|z-z_1|^\psi}\right)
N_{x,y_1,z_1},
\label{(2.15)}
\end{align}
(Recall that $G_\lambda^{(k)}$ contains an identity matrix in the block $R^{(k)}$. This leads to the Kronecker $\delta$ terms in (\ref{(2.15)}), as we need to consider separately the cases $y \in B_{x,k}$, $y \notin B_{x,k}$, and $z \in B_{x,k}$, $z \notin B_{x,k}$.) Taking the expectation and inserting (\ref{(2.17)}), we obtain decay from $y$ to $z$ through the intermediate points $y_1,x,z_1$. Thus
\begin{equation}
\mathbb{E}\,|\varphi_\alpha(y)\varphi_\alpha(z)| \le 4\veps^{q'|y-z|^\chi}.
\label{(2.18)}
\end{equation}
for some $q'<q/8$ (recall that $\chi = \psi/3$: the probability decay is slower than the graphical decay). Replacing $\veps$ with $\gamma^\phi$, we obtain the corresponding bound in (\ref{(2.4a)}). 
This completes the proof of Theorem \ref{thm:2.2}.\qed

We may prove a refined version of Theorem \ref{thm:2.2} in which only spectrum in an interval $[E-\delta/2,E+\delta/2]$ is considered. The resulting bound combines the density of states estimate from Theorem \ref{thm:2.1} with the decay estimate just completed.
\begin{corollary}
\label{cor:2.2'}
Let $\chi = \tfrac{2}{9}$. There exists a constant $\bar{q}>0$ such that for $\gamma$ sufficiently small, the bound 
\begin{equation}
\mathbb{E} \sum_{\alpha:\,\lambda_\alpha \in [E-\delta/2,E+\delta/2]} |\varphi_\alpha(y)\varphi_\alpha(z)| \le \gamma^{\bar{q}(|y-z|\vee L_{\hat{k}-2})^\chi}.
\label{(2.18a)}
\end{equation}
holds for any $\delta \le \veps/2$ and any rectangle $\Lambda$. Here $\hat{k}$ is defined through the inequality $\veps_{\hat{k}+1} < 2\delta \le \veps_{\hat{k}}$.
\end{corollary}
\textit{Proof}. Adopting a hybrid approach, we take $E$ to be the central energy in the Schur
complement up through step $\hat{k}$. As in the proof of Theorem \ref{thm:2.1}, we find that the probability that $x$ is in $R^{(\hat{k})}$ is $\le \veps^{qL_{\hat{k}-2}^\chi}$, \textit{c.f.} (\ref{(2.3)}).
Then we initiate the energy-following procedure, picking $x \in R^{(\hat{k})}$ and the corresponding block $B_{x,\hat{k}}$ of $R^{(\hat{k})}$. We find a solution to
$\lambda \in \text{spec}\,\tilde{F}_\lambda^{\hat{k}}(B_{x,\hat{k}})$ in $[E-\veps_{\hat{k}}/3,
E+\veps_{\hat{k}}/3]$, and continue as before.  Now when the four cases are considered, 
\textit{c.f.} (\ref{(2.10)}), there is a minimum decay length $L_{\hat{k}-2}$, and a corresponding bound
\begin{equation}
\mathbb{E} \, N_{x,y}^{(i)} \le \veps^{q(|x-y|\vee L_{\hat{k}-2})^\chi/3}, \text{ for }i=1,\ldots,4.
\label{(2.18b)}
\end{equation}
If we consider (\ref{(2.11)}), we have $\underline{k}\ge\hat{k}$, as all blocks are resonant on scale $\hat{k}$. In (\ref{(2.12a)}), we have $\tilde{k}\ge\hat{k}$ and $\hat{k}\le j\le \tilde{k}$, which also guarantees a decay of the form (\ref{(2.18b)}). Likewise, in
(\ref{(2.12b)}),(\ref{(2.12c)}), we have $\tilde{k}\ge\hat{k}$, and a similar estimate holds. This completes the proof.\qed

\subsection{Level Spacing}\label{ssec:level}
Let us now state our main result on minimum level spacing for $H$. Let $\{E_\alpha\}_{ \alpha = 1,\ldots,2|\Lambda|}$ denote the eigenvalues of $H$.
\begin{theorem}
\label{thm:2.3}
There exists a constant $b>0$ such that for $\gamma$ sufficiently small, 
\begin{equation}
P\Big(\min_{\alpha \ne \beta} |E_\alpha  - E_\beta| < \delta\Big) \le |\Lambda| \exp\left(-b|\log \gamma|^{3/4}|\log \delta|^{1/4}\right),
\label{(2.19)}
\end{equation}
for any rectangle $|\Lambda|$ and any $0<\delta\le \gamma^{1/4}/4$.
\end{theorem}
This is a theorem about microscopic level statistics: we are interested in the behavior as $\delta \rightarrow 0$. We see that the probability goes to zero faster than any power of $1/|\log \delta|$.

\textit{Proof.} We modify the argument in (\ref{(2.11)})-(\ref{(2.14)}) in the previous proof. Let us define $N_x(\delta)$ to be the number of eigenvalues $\lambda$ of $H$ that can be reached via the energy-following procedure starting at $x$, and which have another eigenvalue within $\delta$ of $\lambda$. Then noting that
\begin{equation}
P\Big(\min_{\alpha \ne \beta} |E_\alpha  - E_\beta| < \delta\Big) \le \sum_x \mathbb{E}\,N_x(\delta),
\label{(2.26)}
\end{equation}
we see that a bound of the form
\begin{equation}
\mathbb{E}\,N_x(\delta) \le \exp\left(-b|\log \gamma|^{3/4}|\log \delta|^{1/4}\right)
\label{(2.27)}
\end{equation}
implies the theorem.

Let us define $k$ by the inequality
\begin{equation}
\veps_{k+1} < 4\delta \le \veps_k.
\label{(2.28)}
\end{equation}
For simplicity we are assuming that $\delta \le \veps/4$ so that (\ref{(2.28)}) holds for some $k \ge 1$. This is only a new restriction for moderately small $|\Lambda|$, in any event, since the bound (\ref{(2.19)}) is nontrivial only for
$\delta \le \exp\left(-b^{-4}\left| \log |\Lambda| \right|^4|/|\log \gamma|^3\right)$. 

As in the proof of Theorem \ref{thm:2.2}, the energy-following procedure generates a sum of terms -- \textit{c.f.} (\ref{(2.11)}),(\ref{(2.12a)}),(\ref{(2.12b)}),(\ref{(2.12c)}) -- that can be used to provide a bound on the expected number of eigenvalues $\lambda$ satisfying some condition. Before, we had the condition that the resonant region includes $x,y$. Here, we have the condition that another eigenvalue
is within $\veps_k/4$ of $\lambda$. Let us break up the sum of terms into part A (for which $\underline{k} > k$) and part B (for which $\underline{k} \le k$). (Recall that $\underline{k}$ is the ``stopping scale,'' \textit{i.e.} the first scale at which
conditions 1-4 all hold.)

Consider first the part A sums. Our previous bounds on terms with $\underline{k} > k$ are already small enough for our current purposes. Thus for the part A sums, we may ignore the condition that defines $N_x(\veps_k/4)$, 
\textit{i.e.} that another eigenvalue is within $\veps_k/4$ of the eigenvalue that is
reached in the energy-following procedure. Taking $y=x$ in each of (\ref{(2.11)}),(\ref{(2.12a)}),(\ref{(2.12b)}), (\ref{(2.12c)}), and restricting to $\underline{k} > k$ or $\tilde{k} = \underline{k}-1 \ge k$, observe that in each case a bound is proven that indicates
rapid convergence in $k$. 
Noting that $\chi = \psi/3$, the bound is $\veps^{qL_k^{\psi/4}/3}$ or better in each case
(times a prefactor $k!$ or $\exp(k^2))$.
We may absorb the prefactor with a decrease in the coefficient in the exponent, and then we obtain a bound $\veps^{qL_k^{\psi/4}/4}$ for the sum of all four cases.

Consider now part B terms, with $\underline{k} \le k$. For such terms, all of 
conditions 1-4 hold at step $k$. In particular, $B_{x,k}$ has reached its maximum
extent, and $\tilde{F}^{(k)}_{E_k}(B_{x,k})$ has a minimum eigenvalue spacing
$\ge \veps_k$ -- see Definition \ref{def:A}. As explained in the proof of Theorem
\ref{thm:2.2}, there is a unique sequence $E_{k+1},E_{k+2},\ldots$ that can
follow $E_k$ in the energy-following procedure. Let $\lambda_0$ be the
limiting eigenvalue. Then as explained above, $|E_k - \lambda_0| \le \veps_k^3$. 
We have the condition that another eigenvalue $\lambda '$ of $H$ is within
$\veps_k/4$ of $\lambda_0$. If that is the case, then we have
$\lambda_0 \in \text{spec}\,F^{(k)}_{\lambda_0}$, $\lambda ' \in \text{spec}\,F^{(k)}_{\lambda'}$ with $|\lambda_0 - \lambda '| \le \veps_k/4$.
We claim that there is some $\tilde{\lambda}'$ with $|\tilde{\lambda}'-E_k| \le \veps_k/3 $ and some level $k$ block $B'$ such that $\tilde{\lambda}' \in \text{spec}\,\tilde{F}^{(k)}_{E_k}(B')$. This follows from making shifts
$F^{(k)}_{\lambda'} \rightarrow F^{(k)}_{E_k} \rightarrow \oplus_B \tilde{F}^{(k)}_{E_k}(B)$ and applying
\begin{align}
\label{eqn:shift}
\|F^{(k)}_{\lambda'} - F^{(k)}_{E_k}\| &\le c_d \frac{\gamma}{\veps}|\lambda ' - E_k|
\le c_d \frac{\gamma}{\veps}\left(\veps_k/4 + \veps_k^3\right), \nonumber\\
\|F^{(k)}_{E_k} - \oplus_B \tilde{F}^{(k)}_{E_k}(B)\| &\le \veps_k^3,
\end{align}
which hold by Theorem \ref{thm:3} and Corollary \ref{cor:2'}. Now $B'$
cannot be $B_{x,k}$, because $\tilde{F}^{(k)}_{E_k}(B_{x,k})$ has minimum
level spacing $\veps_k$, and $E_k \in \text{spec}\,\tilde{F}^{(k)}_{E_k}(B_{x,k})$.

The probability that both $B_{x,k}$ and $B'$ are
blocks of $R^{(k)}$ is bounded by $\hat{P}^{(k)}(B_{x,k})\tilde{P}^{(k)}(B')$. 
(By construction, the probabilities $(3\sqrt{\veps})^{n(B)}$ and $\veps^{L_k^{3\psi/5}/3}$ in (\ref{(1.30)}),(\ref{(2.6a)}) correspond to independent events that must 
occur if $B_{x,k}$ and $B'$ are both blocks of $R^{(k)}$.)
So we may sum $\tilde{P}^{(k)}(B')$ over all possibilities for $B'$ for each
$B_{x,k}$ that arises in the energy-following procedure. This is bounded
by the sum over $y$ of the probability that $y \in R^{(k)} \setminus B_{x,k}$.
By (\ref{(2.3)}), this is bounded by $|\Lambda|\veps^{qL_{k-2}^\chi}$. The net
result is that the sum of part B terms for $\mathbb{E}N_x(\veps_k/4)$ is
bounded by the same four sums (\ref{(2.11)}),(\ref{(2.12a)}),(\ref{(2.12b)}),(\ref{(2.12c)}), times an overall factor of $|\Lambda|\veps^{qL_{k-2}^\chi}$.
Taking $x=y$, we have by (\ref{(2.14)}) that the combined sums are bounded by 3,
and this leads to an overall bound of 
$3|\Lambda|\veps^{qL_{k-2}^\chi}$ on the sum of part B terms.

Adding the above bounds for parts A and B, we obtain
\begin{equation}
\mathbb{E}\,N_x(\delta) \le \veps^{qL_k^{\psi/4}/4} + 3|\Lambda|\veps^{qL_{k-2}^{\psi/3}}.
\label{(2.29)}
\end{equation}
Let us make a provisional assumption that $|\Lambda| \le \veps^{-qL_k^{\psi/4}/5}$. Then 
we have
\begin{align}
\mathbb{E}\,N_x(\delta) &\le \veps^{qL_k^{\psi/4}/4} + 3\veps^{q\left[L_{k}^{\psi/3} 2^{-2\psi/3} - L_k^{\psi/4}/5\right]}
\nonumber \\
&\le \veps^{qL_k^{\psi/4}/4} + 3\veps^{qL_{k}^{\psi/3}/4}
\nonumber \\
&\le 4\veps^{qL_k^{\psi/4}/4} \le \veps^{qL_{k}^{\psi/4}/5},
\label{(2.30)}
\end{align}
using $2^{-2\psi/3} - \tfrac{1}{5} \ge \tfrac{1}{4}$. Summing over $x \in \Lambda$, we obtain that
\begin{equation}
P\big(\min_{\alpha \ne \beta} |E_\alpha  - E_\beta| < \veps_k/4\big) \le |\Lambda| 
\veps^{qL_{k}^{\psi/4}/5}.
\label{(2.31)}
\end{equation}
The provisional assumption $|\Lambda| \le \veps^{-qL_k^{\psi/4}/5}$ is valid in our proof of
(\ref{(2.31)}), since if it is false, (\ref{(2.31)}) is automatically true.
Recalling that $\veps_k = \veps^{L_k^\psi}$, we may write
\begin{equation}
\veps^{qL_{k}^{\psi/4}/5} = e^{-(q/5)|\log \veps| |\log_\veps \veps_k|^{1/4}} = 
e^{-(q/5)|\log \veps|^{3/4}|\log\veps_k|^{1/4}}.
\label{(2.32)}
\end{equation}
Then since $\delta > \veps_{k+1}/4 = \veps_k^{2^\psi}/4$, we have a bound $|\log \delta| \le 2|\log \veps_k|$
for $\veps = \gamma^\phi$ small. Thus after inserting (\ref{(2.32)}) into (\ref{(2.31)}), we 
obtain the statement of Theorem \ref{thm:2.3}.\qed

\textit{Remark 3}. The second term on the right-hand side of (\ref{(2.29)}) is subdominant for small
$\delta$, as shown above. If we were to keep it explicit, it would lead to a term proportional to
$|\Lambda|^2$, corresponding to cases where two separate blocks are resonant to each other. If one wishes to obtain a level-spacing bound for $\gamma^{1/4}/4 < \delta \le 1$, that term becomes the dominant one. To see how this affects the level-spacing estimate, look directly at the level-spacing condition for $\{t_x\}_{x\in \Lambda}$, and apply (\ref{(1.11)}):
\begin{equation}
P\big(\min_{x \ne y} |t_x - t_y| < 2\delta\big) \le \tfrac{1}{2}|\Lambda| (|\Lambda| - 1) \cdot 2\sqrt{2\delta} \le \sqrt{2\delta}|\Lambda|^2.
\label{(2.32a)}
\end{equation}
The eigenvalues of $H$ differ from $\{\pm t_x\}_{x\in \Lambda}$ by $O(\gamma)$, by Weyl's inequality. Hence
\begin{equation}
P\big(\min_{\alpha \ne \beta} |E_\alpha  - E_\beta| < \delta \big) \le \sqrt{2\delta} |\Lambda| ^2, \text{ for } \gamma^{1/4}/4 < \delta \le 1.
\label{(2.32b)}
\end{equation}

\textit{Remark 4}. We see a decrease $L_k^{\psi/3}\rightarrow L_k^{\psi/4}$ in the exponent of the
probability bounds when switching from density of states estimates to level spacing estimates. This arises because the discriminant is a polynomial of degree $4n^2$, 
whereas the determinant has degree $n$. With some optimization of our procedure, 
one should be able to obtain exponents close to $L_k^{\psi/2}$, $L_k^{\psi/3}$, 
respectively, with $\psi$ close to 1. These correspond to the worst case in a tradeoff between $\veps^n$ and $\veps^{L_k^\psi/n}$ or $\veps^{L_k^\psi/n^2}$, respectively.

\begin{footnotesize}
\bibliographystyle{acm}

\end{footnotesize}
\end{document}